\documentclass[11pt]{article}
\usepackage[margin=1in]{geometry}

\usepackage[dvipsnames]{xcolor}
 
\usepackage{amsmath}
\usepackage{mathtools}
\usepackage{physics}
\usepackage{braket}
\usepackage{dsfont}
\usepackage{relsize}
\usepackage{amsfonts}
\usepackage{amssymb}
\usepackage{mathrsfs}
\usepackage{cancel}

\usepackage{cite}

\usepackage{caption}
\usepackage{rotating} 

\usepackage{subcaption}
\usepackage{graphicx} 
\graphicspath{ {./images/} }

\usepackage[rightcaption]{sidecap}

\usepackage{wrapfig}

\usepackage{bbold}
\usepackage{amsthm}
\newtheorem{thm}{Theorem}
\newtheorem{lemma}{Lemma}

\usepackage{float}
\usepackage{cleveref}
\usepackage{relsize}
\usepackage{amsmath,amssymb,amsthm,eucal,mathrsfs,yfonts,bbm,bbm}
\usepackage[T1]{fontenc}
\usepackage{tikz}

\usetikzlibrary{calc, positioning, shapes, through, intersections}

\usepackage{chngcntr}
\counterwithin{figure}{section}

\newcommand{\hscalar}[2]{\langle #1 , \, #2 \rangle }

\newcommand{\nn}{\nonumber}

\newcommand{\pa}{\partial}

\numberwithin{equation}{section}

\title{\textbf{Quantum backflow in the presence of a purely transmitting defect}}

\author{Alexandre Hefren de Vasconcelos Jr\footnote{alexandre.hefren@york.ac.uk}}
\date{%
    \textit{  Department of Mathematics,\\%
      University of York, \\
       York YO10 5DD, \\
       United Kingdom\\}
}

\begin{document}

\maketitle

\begin{abstract}
We analyse the quantum backflow effect and extend it, as a limiting constraint to its spatial extent, for scattering situations in the presence of a purely transmitting discontinuous jump-defect. Analytical and numerical comparisons are made with a different situation in which a defect is represented by a  $\delta$ function potential. Furthermore, we make the analysis compatible with conservation laws. 
\end{abstract}

\newpage

\section{Introduction}
Quantum theory certainly has different mathematical formulations and significant conceptual differences that go beyond a relativistic extension of quantum mechanics to quantum field theory \cite{Bert}. Nevertheless, quantum theory shares some basic ideas such as the Heisenberg's uncertainty principle among any of its formulations or extensions. Effects related to the uncertainty principle may arise as inequalities. For example, the ``quantum inequalities'' in quantum field theory \cite{Ford1978,FewsterQFT}, which are lower bound restrictions on the fluxes and energy densities of physical systems, and the quantum backflow phenomenon \cite{Allcock} for the probability current in quantum mechanics.

Similarly to the quantum energy inequalities, which are limitations on the magnitude and duration of negative energy densities, the backflow inequality can be stated in its time-averaged or spatial-averaged versions. The total energy of a physical system being bounded below is a fact related to the existence of a stable ground state. Nonetheless, there is an incompatibility between positive energy density conditions and local quantum fields \cite{EpsteinGlaser}. The lower bound on the backflow effect, however, does not seen to have an immediately clear physical interpretation. Consideration of both effects in a common framework such as a free relativistic theory may provide some insight on their relationship. In fact, whilst most of the work on quantum backflow considered only the situation without any internal degree of freedom and non-relativistic, the case for a free Dirac particle was studied in \cite{Bracken2}, for instance. Moreover, as the energy is usually considered in connection with a conservation law, it is reasonable to do the same for the backflow analysis and associate a conservation law with it when possible. That is not possible for an interaction described by a $\delta$ potential function, but a jump-defect provides us with this possibility.

Interaction-free situations present a playground for numerous discussions, but more realistically one has to consider the effect of interaction. In \cite{Henning}, the backflow effect was extended to scattering situations in short-range potentials. It reinforced the universality of quantum backflow beyond a free theory and also stated that the lower bound feature, the constraint on how negative it can be, is stable under the inclusion of interaction. Although their work has proved the existence of lower bound estimates for a particular class of short-range potentials, they also noticed that a very  short-range $\delta$ potential, although outside their theorem's validity, has a limited backflow effect. A special particularity of the $\delta$ is that it can be seen as a potential function, but it can also be seen as a point-defect that is characterised by some sewing conditions at the defect's location. Knowing that, we ask ourselves about the possibility of including other kinds of point-defect described by sets of sewing conditions in the discussion of the quantum backflow effect. Defects were previously considered in scattering situations \cite{Delfino}, and integrable defects are generally categorised as purely transmitting \cite{Konik1999}. In particular, we consider a jump-defect that is purely transmitting in the context of non-relativistic quantum mechanics in one spatial dimension. In this sense, the jump-defect is similar to the P\"{o}schl-Teller potential \cite{Teller1933} given by 
\begin{equation} 
V(x) = -\frac{\mu(\mu + 1)}{2\cosh^2{x}}, \quad \mu > 0 .
\end{equation}
However, the latter is only reflectionless when the parameter $\mu$ is taken to be an integer while the jump-defect is always purely transmitting.

The jump-defect is halfway between the P\"{o}schl-Teller and the $\delta$ potential, but there are two relevant features that make it very different from the $\delta$. Because it is purely transmitting, all contributions towards the negative probability fluxes come solely from the superposition of positive momentum states rather than a mixture of scattering and the superposition of positive momentum states. It also allows us to keep conserved quantities that were conserved in the free case, such as energy, momentum (probability flux) and probability. As a point-defect, both of them involve some kind of discontinuity. But while the $\delta$ has a discontinuous first derivative of the wavefunction, the jump-defect has a discontinuous wavefunction. Specifically, the wavefunction discontinuity requires one to work with distinct multiple domains rather than a single domain. The present work extends the quantum backflow effect to this situation, in which the interaction is represented by a set of sewing conditions describing a discontinuous transparent jump-defect, and shows that the effect has a finite spatial extent, or a lower bound. It also extends the previous analysis \cite{Henning} for the $\delta$-case by scanning different values of the parameters and unveiling some structure in the attractive case. 

This paper is composed of seven sections and one appendix. 
In sections~\ref{sec:qbackflow} and \ref{sec:interactscatter}, we present the quantum backflow setting for both the free case and in the presence of a scattering potential.
Section~\ref{sec:integrable} focuses attention on a particular defect, namely a discontinuous jump-defect in the linear Schr{\"o}dinger equation.     
Section~\ref{sec:backflowindefect} introduces the calculations for the backflow effect in the presence of the $\delta$-defect and the jump-defect. In section~\ref{sec:results}, numerical details are provided along with the results in two-dimensional plots for both the $\delta$-case and the jump-case. We summarise the work in the concluding remarks in section~\ref{sec:remarks}.
Finally, the Appendix contains three-dimensional plots encapsulating the behaviour of the lowest backflow eigenvalue under changes of the defect parameter and the position of measurement.

\section{Quantum backflow}\label{sec:qbackflow}

In non-relativistic quantum mechanics, the continuity equation for the probability density in one space dimension is
\begin{equation}
\partial_t \rho = - \partial_x j\,,
\end{equation}
where $ \rho = \lvert\psi\rvert^2$ is the probability density, $j$ is the probability current density and $\psi$ the square-integrable wavefunction of the system. The Schr{\"o}dinger equation for the wavefunction of a quantum system is simply
\begin{equation}\label{schrodinger}
i\hbar\partial_t\psi = H\psi\,,
\end{equation}
where $H$ is the Hamiltonian operator associated with the system. The state vector is commonly denoted by $\ket{\psi} \in \mathcal{H}$, as an abstract vector in the Hilbert space
of the physical system. Not all solutions of this equation are elements of the space of (equivalence classes of) square-integrable functions $L^{2}(\mathbb{R})$,
but these solutions are crucial for scattering theory. As a consequence of the Schr{\"o}dinger equation, in the free case, one has
\begin{equation}\label{eq:j_psi}
     j_{\psi}(x)=\frac{i\hbar}{2m}\left(\partial_{x}{\psi^{\star}(x)}\psi(x)-\psi^{\star}(x)\partial_{x}{\psi(x)}\right) := 
     \hscalar{\psi}{J(x)\psi}
\end{equation}
where now the $\psi$-dependence is explicitly indicated, and $j_{\psi}(x)$ can be expressed in terms of the associated quadratic form $J(x)$. The space average of \eqref{eq:j_psi} with a test function, generally $f \in \mathcal{S}(\mathbb{R})$ in Schwartz-class
is given by

\begin{equation}
  j_\psi(f) =   \hscalar{\psi}{J(f)\psi} = \int dx\,f(x)\,j_\psi(x) \,,
\end{equation}
and is understood as the spatial-averaged probability current measured by a spatially extended apparatus. The corresponding smeared operator is the integration $J(f) = \int f(x)J(x)dx,$ understood in the sense of quadratic forms. This operator is Hermitian for a real function $f$ and is written as
\begin{equation}\label{eq:Jf-Schrodinger}
     J(f)=\frac{1}{2m}\left(\hat{P}f(\hat{X})+f(\hat{X})\hat{P}\right),
\end{equation}
with position operator $\hat{X}$ and momentum operator $\hat{P}$.

The effect that, for a particle with positive momentum, the probability of finding it to the right of some reference point may decrease with time is simply called quantum `backflow'\cite{Henning,Fewster,Bracken,Berry,Goussev}. This means that given a wavefunction $\tilde{\psi}$ with support in momentum space restricted by supp $\left(\tilde{\psi}\right) \subset \mathbb{R}_{+}$, right-moving wave function, it is not guaranteed at all that the probability current density fulfills the positivity condition  $j_{\psi}(x) >0$ with $x \in \mathbb{R}$. The backflow effect has been discussed in both temporal-averaged and spatial-averaged versions.

\section{Interaction in scattering situations}\label{sec:interactscatter}

In interaction-free situations, the maximal amount of backflow, spatially averaged with a positive test function $f$, i.e. the lowest bound, is defined \cite{Henning} by
\begin{equation}\label{eq:beta0}
     \beta_0(f):= \inf \expval{ E_+J(f)E_+ }_{\psi}\, ,
\end{equation}
where the infimum is understood as 
$$\inf\expval{A} := \underset{\norm{\psi}=1}{\inf} \hscalar{\psi}{A\psi} \in \left( -\infty,\infty\right).$$ 
Hence, $\beta_0(f)$ is the minimum eigenvalue of the averaged current evaluated in right-moving states, $E_+J(f)E_+$. The orthogonal projection $E_+$ of the momentum operator makes sure that the momentum is positive ($k>0$). The question of how negative this quantity can be, and if it is actually bounded below, was answered in \cite{Fewster}, where it was proved that a state-vector independent quantum inequality ensures $\beta_0(f)$ is bounded below. While a lower bound is attained, the same is not true for an upper bound, and $E_+J(f)E_+$ is unbounded above, exactly as the corresponding non-smeared version $E_+J(x)E_+$. 

For scattering situations, we consider the effect of an interaction with a potential term $V$, external and time-independent for simplicity, added to the free Hamiltonian so that
\begin{equation} 
H = \frac{{\hat{P}}^2}{2m}+ V(\hat{X}).
\end{equation}
As a physical requirement, the potential is Hermitian. While the concept of right-movers is clear in a free case, the time evolution associated with an interacting Hamiltonian does not commute with the projector $E_+$, meaning that the space of right-movers $E_+L^{2}(\mathbb{R})$ is not invariant under time evolution transformations. As an alternative equivalent concept, we adopt the asymptotic right-movers in the sense of scattering theory, as used before in \cite{Henning}. In this way, we consider a state such that its incoming asymptote is a right-mover.
The incoming M\o{}ller operator is given by
\begin{equation}\label{omega:time}
\Omega^{\textrm{(IN)}} = \Omega_V := \operatorname*{s-lim}_{t \rightarrow - \infty} e^{+iHt} e^{-iH_0 t}\,,
\end{equation}
with $\operatorname{s-lim}$ denoting the strong operator limit and $H_0$ the free Hamiltonian. 
Our quantity of interest is now dependent on the potential and defined as
\begin{equation}\label{main}
\beta_V(f) := \inf\expval{ E_+ \Omega_V^{\dagger} J(f) \Omega_V E_+}_{\psi}\,,
\end{equation} 
which is called the ``asymptotic backflow constant'' \cite{Henning} and it is the lowest eigenvalue of the operator $E_+ \Omega_V^{\dagger} J(f) \Omega_V E_+$. In the future, we will refer to that as the ``asymptotic current operator'' or simply the ``interacting current''.

To ensure the existence of the scattering theory, we usually work with potentials that vanish sufficiently fast at spatial infinity. This is based on the fact that the fall-off properties of the potential are related to smoothness properties of the scattering data. Specifically, we require the fulfillment of the condition \cite{Faddeev}

\begin{equation}
  \norm{V}_{1+} := \int_{-\infty}^{+\infty} dx \left(1 +|x|\right) \mid \! V(x)\! \mid \, < \infty \;, 
\end{equation}
and we say that $V \in L^{1+}(\mathbb{R})$. In the stationary scattering theory, one has the time-independent Schr{\"o}dinger equation (TISE) for a wavefunction $\varphi(x)$ 
\begin{equation}\label{TIschrodinger}
\left(-\frac{\hbar^2}{2m}\frac{\partial^2}{\partial x^2}+ V(x)\right)\varphi(x) = \frac{(\hbar k)^2}{2m}\varphi(x),
\end{equation}
for which the scattering solutions $x \to \varphi_k(x)$ with $k>0$ have asymptotics of the form
\begin{equation}\label{eq:asymp}
\varphi_k(x) = \begin{cases} 
T_V(k) e^{ikx} + o(1)  & \text{as } x \rightarrow \infty, \\
 e^{ikx} +R_V(k)e^{-ikx} + o(1) & \text{as }x \rightarrow -\infty\,,
\end{cases}
\end{equation} 
with transmission $T_V(k)$ and reflection $R_V(k)$ coefficients. 

Either working with perturbation approximations or the exact solution, a key ingredient for analysing the backflow effect in scattering situations is the expansion of the M\o{}ller wave operator in the following integral form; see, for example, \cite{Yafaev:analytic} for the Lemma below. 
\begin{lemma}\label{mollerexpansion}
Let $V\in L^{1+}(\mathbb{R})$. Then the operator $\Omega_V$ defined in \eqref{omega:time} exists. Further, the solution $x \mapsto \varphi_k(x)$ ($k>0$) of \eqref{TIschrodinger} with the asymptotics \eqref{eq:asymp} exists and is unique, and for any $\tilde\psi \in C_0^\infty(\mathbb{R})$, 
\begin{equation}\label{eq:omegakernel}
\bra{x}\Omega_VE_+ \ket{\psi} = 
(\Omega_V E_+ \psi)(x) = \frac{1}{\sqrt{2\pi}}\int_0^\infty dk\, \varphi_k(x) \tilde\psi(k)\,.
\end{equation} 
\end{lemma}
\noindent
By the use of some estimates, e.g., \cite{Faddeev,Deift}, the following theorem \cite{Henning} is a result on the existence of backflow in scattering situations and also on its lower bound.
\begin{thm}
Let the potential function $V$ be a $L^{1+}(\mathbb{R})$-class potential, i.e. \\
$\norm{V}_{1+} < \infty$ , for any $f \in \mathcal{C}^{\infty}_{c}(\mathbb{R})$, with $ f \geqslant 0$, $\exists \; C_{V,f} > 0$ such that 
\begin{equation}
\expval{E_{+}\Omega_V^{\dagger}J(f)\Omega_{V}E_{+}}{\psi} \geqslant -C_{V,f} \quad \text{for} \;  \norm{\psi}=1.
\end{equation}
\end{thm}
\noindent
Hence, the asymptotic backflow constant is finite, $\beta_V(f) \!>\! -\infty$.
The existence of backflow and the boundness (below) of backflow are stable under the addition of a scattering potential to the Hamiltonian. This means that, even in the presence of reflection, the effect is bounded below. Henceforth, we denote the expectation value of the interacting operator
in a general (normalized) state vector $\ket{\psi}$ by 
\begin{equation}\label{eq:interactingcurrent}
\expval{J_V(f)}_{\psi} \coloneqq \bra{\psi}{ E_+ \Omega_V^{\dagger} J(f) \Omega_V E_+}\ket{\psi}.
\end{equation}
Moreover, the expansion of this expectation value relies on the use\footnote{The Lemma~\ref{mollerexpansion} requires $\tilde{{\psi}}$ to be smooth of compact support $\mathcal{C}^{\infty}_{c}$. However, $\mathcal{C}^{\infty}_{c}(\mathbb{R})$ is dense in $L^2(\mathbb{R})$ and through the use of Friedrichs extensions \cite{ReedSimon} the discussion applies to a general $\psi$ in the domain $D(J) $ of our operator.} of the Lemma~\ref{mollerexpansion}.

As is the case for the Hamiltonian, we expect that the asymptotic current operator has a spectrum composed of pure point and absolutely continuous parts. Thus, we have some eigenvalues, with the lowest one denoted by $\beta_V(f)$, and at some point a continuum of ``generalized'' eigenvalues. It is important to stress our interest in this lowest eigenvalue in the context of quantum inequalities.

\section{Integrable defects}\label{sec:integrable}
Either in classical or quantum theory, partial differential equations come in to describe the dynamics of the systems we want to study. The same physical idea can be implemented in different ways, depending on what one wants to describe. Integrable defects \cite{Delfino,Konik1999,Mintchev2002,Ed2003,Ed2006,Caudrelier2008,Ed2009} can be treated both in classical and quantum contexts in linear and nonlinear theories. In a linear theory, integrability is certainly redundant, but the underlying motivation is the same: to preserve conservation laws. 

In the Schr\"{o}dinger equation for a wavefunction $\varphi$, one explicitly writes down a potential term, usually a function of the position, in the Hamiltonian of the system. In case the potential is only a function of the space coordinate (and possibly of time), but not of $\varphi$ itself, the equation is still a linear partial differential equation. Additionally to working with an explicit potential term, there is another way of implementing interactions in the presence of point-like impurities or defects, a kind of internal boundary at a point. Rather than written as an external potential function, the defect can be described by a set of sewing conditions. In $1+1$ dimensions, these conditions relate the field and its derivatives on the left to the field and its derivatives on the right of the defect's location.  

The $\delta$-type defect has the pedagogical advantage of allowing both descriptions; it can be written as the usual delta potential $\delta(x)$ or as a set of two sewing conditions. In particular, for the $\delta$-type defect, one condition is a statement of the continuity of the field at the defect location and the other one describes the discontinuity of the spatial derivative of the field. Although interesting and more familiar, the $\delta$-type of impurity may spoil the integrability of a nonlinear integrable system. For instance, that is the case for the sine-Gordon equation \cite{Goodman}. However, some years ago it was shown that there exist two types of defects that are integrable, proved by constructing Lax pairs, and they were categorised as type I and type II \cite{Ed2009}. The former is simpler in the sense that only the field has dynamics, and the latter is a generalization with an extra function defined on the defect; it has an extra internal degree of freedom.

In this work, we focus on the type I integrable defects. While the $\delta$-type defect has continuous solutions at the defect location, we can have a defect that allows a discontinuity of the field at the same location. Such a defect, with a particular set of sewing conditions, is called a ``jump-defect''. In the context of fluid mechanics, such defects are very similar to shock waves, for example, which have sewing conditions expressed by the Rankine-Hugoniot conditions \cite{Rankine}.

\begin{subsection}{Jump-defect in non-relativistic context}
Although a Lagrangian description is not the only way for setting up the situation we are interested in, we can start from a Lagrangian in $1+1$ dimensions. 
By conveniently setting a length scale $\ell$ as the unit of length and $m\ell^2/\hbar$ as the unit of time, equivalently setting $\hbar= m =1$, the Lagrangian density is
\begin{equation}\label{LSLagrangian}
    {\cal L}[\psi] = \frac{i}{2}\left(\psi^{\star} \psi_t - \psi^{\star}_t \psi\right) - \frac{|\psi_x|^2}{2} \, ,
\end{equation}
and the Euler-Lagrange equation gives the linear Schr\"{o}dinger equation
\begin{equation}\label{LS}
    2i \psi_t + \psi_{xx} =0 .
 \end{equation}

The defect can be placed at the position $x_D =0$ on the real line, for example. This means that the bulk region, $-\infty < x < \infty$, will effectively split in two parts, as in figure~\ref{fig:realline}. The field on the left of the defect ($x<0$) will be denoted $u=u(x,t)$ and the field on the right ($x>0$) will be denoted $v=v(x,t)$.
\begin{figure}
\centering{
\begin{tikzpicture}

\draw[line width=0.25mm, black, thick, <-> ] (-4,0) -- (4,0);
  \draw [fill] (0,0) circle [radius=0.030];
  \node [below] at (0,0) {$x_D=0$};
   \node [below] at (-2,0) {u};
  \node [below] at (2,0) {v};
 
 \end{tikzpicture} 
\caption{Locating the defect on the real line} \label{fig:realline} }
\end{figure}
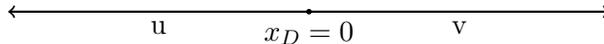 
From the nonlinear Schr\"{o}dinger model considered in \cite{Ed2006}, we particularise to the linear case, where $u$ and $v$ obey the linear Schr\"{o}dinger equation, with a Lagrangian density composed of three contributions coming from $u$, $v$ and the defect
\begin{equation}
\label{lagrangian} {\cal L}=\theta(x_D-x){\cal L}[u] +
\theta(x-x_D){\cal L}[v] +\delta (x-x_D) B[u,v]\, ,
\end{equation} 
where the Heaviside function is $\theta(x) = 0$ for $x < 0$ or $\theta(x)=1$ for $x>0$ and
\begin{equation}\label{NLSB}
    B[u,v]=\alpha\, \left[\frac{i}{4\alpha^2} \left( (u-v)(u^{\star} - v^{\star})_t - (u-v)_t(u^{\star}- v^{\star}) \right) + \frac{1}{4}(u+v)(u^{\star} + v^{\star}) \right] 
\end{equation}
with a real parameter $\alpha$.
The defect conditions follow from the variation of the Lagrangian and are given by
\begin{eqnarray} \nn \label{NLSlinear}
 u_x&=&-\frac{i}{\alpha}(u - v)_t +\frac{\alpha}{2}(u+v)\; , 
 \\
v_x&=&-\frac{i}{\alpha}(u - v)_t -\frac{\alpha}{2}(u+v)\;, 
\end{eqnarray}
which can immediately be rearranged as
\begin{eqnarray} \nn \label{sewing}
    u_x - v_x &=& \alpha(u+v)\; ,\\
    v_x + u_x &=& -\frac{2i}{\alpha}(u-v)_t 
    \; ,
\end{eqnarray}
both valid at the defect's position $x = x_D =0$. The first thing to notice is that the difference of the spatial derivatives is proportional to the average (arithmetic mean) value of the fields meeting at the defect's location, and the parameter $\alpha$ works as a strength of that difference. The second is the discontinuity at the defect, namely we can have $u \neq v$. For reference, we shall mention that in the $\delta$ potential function case, with potential $V(x) = \lambda \delta(x)$, the conditions are 
\begin{equation}\label{deltasewing}
    u=v, \quad  (v_x - u_x) = 2\lambda u,  
\end{equation} 
both evaluated at the defect's position $x = x_D =0$, and $\lambda$ is the associated defect parameter. Note also the similarity to the B\"{a}cklund transformations \cite{Lamb1974} for the linear Sch\"{o}dinger equation. 
It would actually be one if the relations were valid for all positions instead of being frozen at the defect's location. 
At the moment, it is unknown if there is a physical system (approximately) described by these sewing conditions \eqref{NLSlinear}. However, they have a physical motivation based on energy, momentum and probability conservation, which will be discussed in the next section, and on the fact that discontinuities are ubiquitous in natural processes. 

The sewing conditions \eqref{NLSlinear} allow the following pair of traveling wave solutions \cite{Ed2006}
\begin{equation}\label{travellingwave}
    u=u_0 \exp(-i \omega t + i kx), \qquad v=v_0 \exp(-i \omega t +i kx) ,\qquad v_0=
    \frac{k+i\alpha}{k-i\alpha}\, u_0,
\end{equation}
where $k$ is real, and the frequency $\omega = k^2/2$ obeys the usual quadratic dispersion relation from the non-relativistic theory.

As our analysis will be restricted to the linear case, we directly look at conservation laws as our guiding principle for the construction of the jump-defect. Specifically, the jump-defect is designed in order to keep valid some conservation laws that are true in the free case. In other words, we ask that the implementation of the jump-defect does not cause a breakdown of the conservation laws we have in the free Schr\"{o}dinger theory.

\end{subsection}

\subsection{Conservation laws}\label{subsec:conservationlaws}

In the free Schr\"{o}dinger case, we know that quantities such as energy, probability and momentum are conserved. For that, we check how a point-defect may affect these conservation laws, and how (if possible) the sewing conditions can modify the quantity so that it remains conserved in the presence of the defect. In particular, we compare the $\delta$-defect to the jump-defect and analyse energy, momentum and total probability. 

Taking only the $u$ contribution, the energy density derived from the Lagrangian density given by \eqref{LSLagrangian} is
\begin{equation}\label{LSenergy}
{\cal E}=\frac{|u_x|^2}{2},
\end{equation}
and similarly for $v$. The total energy, either calculated as the expectation value of the Hamiltonian or as the spatial integral of the energy density, has to be split up into contributions from two domains when we have one single point-defect. Each domain is separately described by a Schr\"{o}dinger equation, and the defect glues together these two regions by the sewing conditions at the defect's location. In fact, for cases where the wavefunction is continuous over the whole domain, it does not make a difference, but we want to include the jump-defect, and discontinuities make a difference. The same applies for all other physical quantities of interest, such as the total momentum and total probability as they are made out of the fields $u$ and $v$.
The total energy is therefore
\begin{equation}\label{Totalenergy}
E =  \frac{1}{2}\int_{-\infty}^{0}  u^{\star}_x u_x dx + \frac{1}{2}\int_{0}^{\infty}  v^{\star}_x v_x dx,
\end{equation}
where we have split up the integral taking into consideration that the defect is located at the origin $x=0$. For checking conservation, we calculate the time derivative
\begin{equation}\nonumber
\begin{aligned}
E_t &=  \frac{1}{2}\int_{-\infty}^{0} ( u^{\star}_{xt}u_x + u^{\star}_x u_{xt}) dx  + \frac{1}{2}\int_{0}^{\infty} ( v^{\star}_{xt}v_x + v^{\star}_x v_{xt}) dx
\\
&=  \frac{1}{2}\left(u^{\star}_t u_x + u^{\star}_x u_{t} \right)\vert_{x=0} - \frac{1}{2}\left(v^{\star}_t v_x + v^{\star}_x v_{t} \right)\vert_{x=0},
\end{aligned}
\end{equation}
where we used the Schr\"{o}dinger equation and ignored the zero contributions at $\pm$ infinity as usual. For the $\delta$-defect, using \eqref{deltasewing} we obtain
\begin{equation*}
\begin{aligned}
    E_t &= -\lambda u^{\star}_t u + \lambda u^{\star}u_t  \\
    &= \left. -\lambda \frac{\pa}{\pa t}(u^{\star}u)\right|_{x=0} \, ,
    \end{aligned}
\end{equation*}
which may not be zero, but we can guarantee the total energy is conserved by modifying it with an extra contribution. That means the adjusted conserved energy is
\begin{equation}
\left.  E_c := E + \lambda u^{\star}u \right|_{x=0} .
\end{equation}
For the jump-defect, we can check that the choice \eqref{NLSlinear} produces
\begin{equation*}
\begin{aligned}
E_t &=  \frac{1}{2} \left[u^{\star}_t \left(\frac{-i}{\alpha}(u-v)_t + \frac{\alpha}{2}(u+v)\right) + u_t \left(\frac{i}{\alpha}(u^{\star}-v^{\star})_t + \frac{\alpha}{2}(u^{\star} + v^{\star})   \right) \right]_{x=0}\\
& \phantom{mm}- \frac{1}{2}\left[v^{\star}_t \left(\frac{-i}{\alpha}(u-v)_t - \frac{\alpha}{2}(u+v)\right) + v_t \left(\frac{i}{\alpha}(u^{\star}-v^{\star})_t - \frac{\alpha}{2}(u^{\star} + v^{\star})   \right) \right]_{x=0}
\\
&= \phantom{m} \left. \frac{\alpha}{4}\frac{\pa}{\pa t}\left((u+v)(u^{\star}+v^{\star})\right)\right|_{x=0},
\end{aligned}
\end{equation*}
which depends on the parameter $\alpha$ and may not be zero, but we can guarantee the conservation by modifying it with the following redefinition of the energy
\begin{equation}
   \left. E_c := E -\frac{\alpha}{4}\vert u+v \vert^2 \right|_{x=0} \, .
\end{equation}
Hence, the energy $E_c$ is conserved.

Now, let us analyse the momentum. The momentum density associated with $u$ is given by
\begin{equation}\label{eq:momentumdensity}
{\cal P}(u) =\frac{i}{2}\left(u^{\star}_x u -u^{\star}u_x \right),
\end{equation}
so that the total momentum is 
\begin{equation}\label{Totalmomentum}
    P =
    \frac{1}{2}\int_{-\infty}^0  i\left( u^{\star}_x u - u^{\star}u_x\right) dx +\frac{1}{2} \int_0^\infty  
    i\left(  v^{\star}_x v - v^{\star}v_x\right) dx .
\end{equation}
We take the time derivative
\begin{equation*}
\begin{aligned}
P_t =& \, \frac{1}{2}\int_{-\infty}^0 i( u^{\star}_{xt}u + u^{\star}_xu_t -  u^{\star}_t u_x - u^{\star}u_{xt}  ) dx +  \frac{1}{2}\int_0^\infty i ( v^{\star}_{xt}v + v^{\star}_x v_t - v^{\star}_t v_x - v^{\star}v_{xt}  ) dx
\\
=& \phantom{m} \frac{1}{4}\left[ -(2u^{\star}_x u_x) + (u^{\star}u_{xx} + u u^{\star}_{xx}) \right]_{x=0} - \frac{1}{4}\left[ -(2v^{\star}_x v_x) +  (v^{\star}v_{xx} + v v^{\star}_{xx}) \right]_{x=0} \, ,
\end{aligned}
\end{equation*}
where we used the Schr\"{o}dinger equation and ignored the zero contributions at $\pm$ infinity.
For the $\delta$-defect, with \eqref{deltasewing},
\begin{equation*}
\begin{aligned}
P_t =& \, \frac{1}{4} \left[   \left(-2(v^{\star}_x - 2\lambda v^{\star})(v_x - 2\lambda v ) + ( -2iv^{\star}v_t + 2i v v^{\star}_t  ) \right)  + (2v^{\star}_x v_x) -   (v^{\star}v_{xx} + vv^{\star}_{xx}) \right]_{x=0} \\
=& \left. \left(\lambda(vv^{\star})_{x} -2\lambda^2 vv^{\star}   \right)\right|_{x=0}  \, ,
\end{aligned}
\end{equation*}
which cannot be written as a time derivative by the use of the sewing conditions. Hence, we are not able to fix this conservation law without any other extra considerations. The momentum $P$ highlights the difference between the $\delta$ and the jump-defect because the same calculation applied to the jump-defect, using \eqref{NLSlinear}, yields

\begin{equation*}
\begin{aligned}
P_t =& -\frac{1}{4} \left[ 2 \left(+\frac{i}{\alpha}(u^{\star} - v^{\star})_t +\frac{\alpha}{2}(u^{\star} + v^{\star})\right) \left( -\frac{i}{\alpha}(u - v)_t +\frac{\alpha}{2}(u+v) \right) \right. \\ & \left. 
\phantom{m} - \left(u^{\star} \left( -\frac{i}{\alpha}(u - v)_{xt} +\frac{\alpha}{2}(u+v)_{x}    \right) + u \left( +\frac{i}{\alpha}(u^{\star} - v^{\star})_{xt} +\frac{\alpha}{2}(u^{\star} + v^{\star})_{x} \right)        \right) \right]_{x=0} \\ 
& \phantom{m} +  \frac{1}{4} \left[   2 \left(+\frac{i}{\alpha}(u^{\star} - v^{\star})_t -\frac{\alpha}{2}(u^{\star} + v^{\star})\right) \left( -\frac{i}{\alpha}(u - v)_t -\frac{\alpha}{2}(u+v) \right)  \right. \\ & \left.
\phantom{m} - \left(v^{\star} \left( -\frac{i}{\alpha}(u - v)_{xt} -\frac{\alpha}{2}(u+v)_{x}    \right) + v \left( +\frac{i}{\alpha}(u^{\star} - v^{\star})_{xt} -\frac{\alpha}{2}(u^{\star} + v^{\star})_{x} \right)        \right)  \right]_{x=0} \, ,
\end{aligned}
\end{equation*}
which can be simplified to
\begin{equation*}
\begin{aligned}
 P_t 
= \, \, -\frac{i}{2}\left [ \frac{\pa}{\pa t}\left( u^{\star}v - v^{\star} u \right) \right]_{x=0} ,
\end{aligned}
\end{equation*}
which may not be zero, but we can guarantee the conservation if we redefine $P$ in order to take in consideration the contribution at the defect's location by
\begin{equation}\label{Termconservemomentum}
  \left.  P_c := P + \frac{i}{2}(u^{\star}v - vu^{\star})\right|_{x=0}.
\end{equation}
Then, the momentum $P_c$ is conserved.

The probability density for $u$ is given by
\begin{equation}
 {\cal N}(u) = u^{\star}u,
\end{equation}
so that the total probability is
\begin{equation}\label{Totalnumber}
N = \int_{-\infty}^0  {\cal N}(u) dx + \int_0^\infty {\cal N}(v) dx = \int_{-\infty}^0  u^{\star}u dx + \int_0^\infty v^{\star}v dx .
\end{equation}
To examine its conservation, we consider
\begin{equation*}
\begin{aligned}
N_t =& \int_{-\infty}^0 (u^{\star}_t u + u^{\star} u_t) dx + \int_0^\infty  (v^{\star}_t v + v^{\star} v_t) dx 
\\
=& \phantom{m} \frac{1}{2} \left(-iu^{\star}_x u +i u^{\star} u_x \right)\vert_{x=0} - \frac{1}{2}  \left(-iv^{\star}_x v +i v^{\star} v_x \right)\vert_{x=0} \, ,
\end{aligned}
\end{equation*}
where we used the Schr\"{o}dinger equation and ignored the zero contributions at infinities. For the $\delta$-defect, using \eqref{deltasewing},
\begin{equation*}
\begin{aligned}
N_t =& \,  \frac{1}{2}\left(  -iu(v^{\star}_x -\lambda u^{\star})\vert_{x=0} + iu^{\star}(v_x -\lambda u)\vert_{x=0} - \left(-iv^{\star}_x v +i v^{\star} v_x \right)\vert_{x=0} \right) \\
=& \phantom{m} \left. \frac{1}{2}  \left(-i v^{\star}_x (u-v) + i (u^{\star} - v^{\star})v_x ) \right)\right|_{x=0} = 0,
\end{aligned}
\end{equation*}
which means this is automatically conserved. 
For the jump-defect, with the choice \eqref{NLSlinear}
\begin{equation*}
\begin{aligned}
N_t =& \, \frac{1}{2} \left[ \frac{u(u^{\star} - v^{\star})_t  + u^{\star}(u-v)_t)}{\alpha} +\frac{i\alpha}{2}\left(u^{\star}(u+v) -u(u^{\star} + v^{\star}) \right) \right]_{x=0} \\
& \phantom{m} -\frac{1}{2}  \left[ \frac{v(u^{\star} - v^{\star})_t  + v^{\star}(u-v)_t)}{\alpha} -\frac{i\alpha}{2}\left(v^{\star}(u+v) -v(u^{\star} + v^{\star}) \right) \right]_{x=0} \\
=& \phantom{m} \frac{1}{2\alpha}\left[\frac{\pa}{\pa t} ((u-v)(u^{\star} - v^{\star})) \right]_{x=0} ,
\end{aligned}
\end{equation*}
which depends on the parameter $\alpha$ and may not be zero. However, we can guarantee conservation if we redefine $N$ to take into account a contribution at the defect by setting
\begin{equation}
   \left. N_c := N - \frac{1}{2\alpha}\vert u-v\vert^2\right|_{x=0} \, .
\end{equation}
 
We have shown how both the $\delta$ and the jump-defect affect some conservation laws and we have seen how we can fix these conservation laws by redefining quantities with an extra contribution which comes from the defect. However, it is clear that the jump-defect allows conservation of $P$ without any further extra information, but the $\delta$-defect does not. When we treat these defects in the context of quantum mechanics, the momentum $P$ is actually related to the probability current. Moreover, we will see how this extra term associated with the defect affects the calculation of the quantum backflow and how it significantly differs from the $\delta$-defect case. Strikingly interesting, in the jump-case, is that the fixing  term, to restore the conservation of $P$, has a substantial contribution to the lowest backflow eigenvalue.    

\section{Backflow in the presence of a defect}\label{sec:backflowindefect}

Because of its importance and as a stepping stone towards the jump-defect case, we first review the backflow calculation in the presence of a $\delta$-defect \cite{Henning}. Before the actual calculation, we need to set the general structure of our quantities of interest.  

For a general interaction, we can write the expectation value \eqref{eq:interactingcurrent} of the interacting current, in position space, as
\begin{equation}
\bra{\psi}{J_V(f)}\ket{\psi} = \int dx \int dx' (\Omega_V E_+ \psi)^{\star}(x') \left[J(f)(x',x)\right] (\Omega_V E_+ \psi)(x),
 \end{equation}
with the kernel $J(f)(x',x)$ in position space. In order to simplify this equation, we need the expression for $J(f)(x',x)$, which can be obtained starting from $\bra{\psi} { J(f)  }\ket{\psi}$ as
\begin{equation*}
 \bra{\psi} { J(f)  }\ket{\psi} = \frac{1}{2} \bra{\psi} { Pf(X) + f(X)P  }\ket{\psi}  = - \frac{i}{2} \int dy \, \psi^{\star}(y) \left(f(y) \frac{\pa \psi(y)}{\pa y} + \frac{\pa}{\pa y}(f(y)\psi(y)) \right), 
 \end{equation*}
which can also be rewritten as  
 \begin{equation*}
 - \frac{i}{2}  \int dy \, \psi^{\star}(y)  \left[\int dy' \,  f(y) \frac{\pa \delta(y-y')}{\pa y}\psi(y')  + \frac{\pa f(y)}{\pa y}\delta(y-y')\psi(y') + f(y)\frac{\pa \delta(y-y')}{\pa y}\psi(y')   \right],
\end{equation*}
where we have used the trick of rewriting $\psi(y) = \int \delta(y-y')\psi(y')dy'$.
Hence, since $\bra{\psi} {J(f) }\ket{\psi} = \int dy \int dy' \psi^{\star}(y') \left[ J(f)(y',y)\right]\psi(y')$, we obtain
\begin{equation}\label{eq:kerneloperatorcurrent}
    J(f)(y,y') = -\frac{i}{2} \left[2f(y) \frac{\pa \delta(y-y')}{\pa y} + \frac{\pa f(y)}{\pa y}\delta(y-y')  \right].
\end{equation}

In abstract Dirac notation, we can write the general structure of the interacting operator current $J_V(f)$. For that, let us first expand our quantity of interest using \eqref{eq:omegakernel} as

\begin{equation}
    \bra{\psi} {J_V(f) }\ket{\psi} = \int dx \int dx'\frac{1}{(\sqrt{2\pi})^2}\int_{0}^{\infty}dk' \, \tilde{\psi}^{\star}(k')  \varphi^{\star}_{k'}(x') J(f)(x',x) 
     \int_{0}^{\infty} dk \,  \varphi_k(x)
     \tilde{\psi}(k).
\end{equation}
For extracting the operator, we insert Dirac delta distributions $\delta(k'-q')$ and $\delta(k-q)$ as
\begin{equation}
\begin{aligned}
 \bra{\psi} {J_V(f)}\ket{\psi} = \frac{1}{2\pi} \int_{0}^{\infty} dk\int_{0}^{\infty}dk' & \tilde{\psi}^{\star}(k') \left[  \int  dq' \delta(k'-q') \int dx \int dx'    \right. \nonumber \\ & \left.
   \times  \varphi_{q'}^{\star}(x')J(f)(x',x) \left(\int dq \; \delta(q-k) \varphi_q(x)\right)\right] \tilde{\psi}(k), 
    \end{aligned}
 \end{equation}
but the delta functions $\delta(k'-q')= \braket{k'|q'}$ and $\delta(q-k) = \braket{q|k}$ can be ``factored'' out,
and we can isolate the abstract operator by the use of the completeness relation to obtain
\begin{equation}\label{eq:interactingoperatorisolated}
  J_V(f) = E_+ \Omega_V^{\dagger} J(f) \Omega_V E_+  = \frac{1}{2\pi}\left[ E_+  \int dq' \ket{q'} \bra{\varphi_{q'}}J(f) 
  \int dq    \ket{\varphi_{q}}\bra{q}  E_+   \right]. \end{equation}
The expression \eqref{eq:interactingoperatorisolated} is the interacting operator expanded in terms of the interacting state vector and it highlights how $J_V(f)$ differs from the free operator $E_+J(f)E_+$. Having the expression of the linear operator is also the starting point for analytical perturbation theory. For a practical calculation such as the lowest eigenvalue, we will work with a basis of the Hilbert space. 

Our expectation value expression \eqref{eq:interactingcurrent} can, therefore, be written as 
\begin{equation}\label{interactingexpectationvalue}
    \bra{\psi}{ J_V(f)} \ket{\psi} = \frac{1}{2\pi}\int_{0}^{\infty}dk \int_{0}^{\infty}dk' \tilde{\psi}^{\star}(k')\tilde{\psi}(k)\int dx \int dx' \left( \varphi_{k'}^{\star}(x')J(f)(x',x)\varphi_k(x) \right),
\end{equation}
where we will denote the inner integrals by
\begin{equation}\label{eq:innerintegral}
 L(k',k) = \int dx \int dx' \left( \varphi_{k'}^{\star}(x')J(f)(x',x)\varphi_k(x) \right).  
\end{equation}
For the lowest backflow eigenvalue expression, we need to take the minimum of the expression \eqref{interactingexpectationvalue} as
\begin{equation}\label{lowesteigenvalueintegralform}
    \beta_V(f) = \frac{1}{2\pi}\int_{0}^{\infty}dk \int_{0}^{\infty}dk' \mathcal{\tilde{J^*}}(k')\mathcal{\tilde{J}}(k) L(k',k),
\end{equation}
where we assume the existence of the lowest eigenvector $\ket{J_{min}}$ of the operator $J_{V}(f)$, for which the associated wavefunction, in momentum space, is denoted by $\mathcal{\tilde{J}}(k)$. At the present moment, however, an explicit analytical solution for the lowest eigenvector is not known even in the free case \cite{Penz,Halliwell}.

\begin{subsection}{Backflow in the presence of a delta-defect} \label{subsect:deltadefect}
Although the $\delta$ potential function is not a $L^{1+}(\mathbb{R})$-class potential (it is not a locally integrable function), it was shown in \cite{Henning} that one can have a (rough) estimate of the lowest backflow eigenvalue, and the numerical results show that the $\delta$ potential is indeed a special case that also has a lower bound for its $\beta_V(f)$. Here we add to the numerical results and the analytical expression for \eqref{eq:innerintegral} of the work contained in \cite{Henning}. The aims for this are twofold: the lowest backflow eigenvalue displays a different behaviour for defect parameter values $|\lambda|<1$ and the analytical calculation of \eqref{eq:innerintegral}, in the $\delta$-defect case, highlights the differences with respect to the discontinuous jump-defect case.  



Let $\varphi_k$ denote the solution for the TISE in the presence of a $\delta$-defect. We can work with derivatives in the weak sense as both $\varphi_k$ and its derivative $\partial_x\varphi_k$ are both locally integrable functions $\varphi_k  \in L_{\textrm{loc}}^{1}(\mathbb{R})$, $\partial_x\varphi_k  \in L_{\textrm{loc}}^{1}(\mathbb{R})$.
The full time-dependent solution to the Schr\"{o}dinger equation  is denoted by
\begin{equation}\label{deltafullsolution}
   \varphi(x,t) = \! \frac{1}{\sqrt{2}}  \int_{-\infty}^{\infty} dk \, \tilde{g}(k) \exp(-iwt)\varphi_{k}(x) = \! \frac{1}{\sqrt{2}} \int_{-\infty}^{\infty} dk \, \tilde{g}(k) \exp(-iwt) \left(\theta(-x)u_k(x) + \theta(x)v_k(x) \right),
\end{equation}
where $\tilde{g}$ is an arbitrary non-zero smoothly varying function used for producing the wave packet as a proper square-integrable $L^2(\mathbb{R})$-solution. As we established before, we denote the solution at the left of the defect by $u$ and at the right by $v$. The time-independent scattering states in position basis \eqref{eq:asymp}, compatible with the sewing conditions \eqref{deltasewing}, are
\begin{eqnarray}\label{deltatimeindep}
    u_k(x) &=&  \exp(ikx) + \frac{\lambda}{ik - \lambda}\exp(-ikx), \quad  x<0 \nonumber \\
    v_k(x) &=& \left(  \frac{ik}{ik - \lambda} \right)\exp(ikx), \quad x>0 ,
\end{eqnarray}
 where the reflection coefficient $R(k)$ and the transmission coefficient $T(k)$ for the $\delta$-defect are explicitly written. We want to concentrate our attention on the time-independent part $\varphi_k(x)$ composed of \eqref{deltatimeindep} and, for that, the inner integral \eqref{eq:innerintegral} reads 
\begin{equation}
 L(k',k) = \int dx \int dx' \left[  \left( \theta(-x') u^{\star}_{k'} + \theta(x') v^{\star}_{k'} \right)  
 J(f)(x',x) \left( \theta(-x) u_{k} + \theta(x)v_{k} \right) \right].
\end{equation}
Since the expression \eqref{eq:kerneloperatorcurrent} for $J(f)(x',x)$ has a factor of $-i/2$, we absorb it by working with $2iL(k',k)$ instead. Each term is expanded by the insertion of the $J(f)(x',x)$ and simplified after integration. Let us focus only on the spatial integrals, namely the kernel $2iL(k',k)$. There are four contributions which we denote by $uu$,$uv$,$vu$ and $vv$. The first contribution ($uu$) to the kernel $2iL(k',k)$ is
\begin{equation*}
 \int dx \int dx' \theta(-x') u^{\star}_{k'}(x')\left(2f(x')\frac{\pa \delta(x'-x)}{\pa x'} + \frac{\pa f(x')}{\pa x'}\delta(x'-x) \right) \theta(-x) u_{k}(x) .
 \end{equation*}
After integration by parts, noting that $\theta^2(x') =\theta(x')$ and simplifying terms, it becomes the first contribution, written in terms of $R(k)$,
\begin{equation}
\begin{aligned}
 &i(k+k')\int dx' f(x') \theta(-x')\exp(ix'(k-k')) + i(k'-k') \int dx' f(x')\theta(-x')R(k) \exp(-ix'(k+k')) \\
& \phantom{mm}  +i(k-k') \int dx' f(x') \theta(-x')R^{\star}(k')\exp(ix'(k+k')) \\
& \phantom{mm} -i(k+k') \int dx' f(x') \theta(-x')R^{\star}(k')R(k)\exp(-ix'(k-k')).
\end{aligned} 
\end{equation}  
The second contribution ($uv$) to the kernel $2iL(k',k)$ is
\begin{equation*}
\begin{aligned}
 \int dx \int dx' \theta(-x') \left(\exp(-ik'x') + R^{\star}(k')\exp(ik'x') \right) &\left(2f(x')\frac{\pa \delta(x'-x)}{\pa x'} + \frac{\pa f(x')}{\pa x'}\delta(x'-x) \right) \\ &\times \left( \theta(x) T(k)\exp(ikx)  \right).
 \end{aligned}
 \end{equation*}
After integration by parts and simplifying, we can use that $\theta(-x') + \theta(x') = 1$, without any problem at the origin as $\varphi_k$ is continuous, to obtain the second term given by 
\begin{equation}\label{eq:secondcontdelta}
\begin{aligned}
    \int & dx' f(x')  \left(\exp(-ik'x') + R^{\star}(k')\exp(ik'x') \right)T(k)\exp(ikx')\delta(x') \\
    &= \left(\frac{ik'}{ik'+\lambda}  \right)\left(\frac{ik}{ik-\lambda}\right)f(0).
    \end{aligned}
\end{equation}
Now, the third contribution ($vu$) to the kernel $2iL(k',k)$ is
\begin{equation*}
\begin{aligned}
 \int dx \int dx' \theta(x') \left( T^{\star}(k')\right)\exp(-ik'x') &\left(2f(x')\frac{\pa \delta(x'-x)}{\pa x'} + \frac{\pa f(x')}{\pa x'}\delta(x'-x) \right) \\ &\times \left( \theta(-x) \left( \exp(ikx) + R(k)\exp(-ikx)  \right) \right),
 \end{aligned}
 \end{equation*}
and it is simplified to become the third term given by
\begin{equation}
\begin{aligned}
 \int & dx' (-f(x'))T^{\star}(k')\left( \exp(ikx') + R(k)\exp(-ikx')\right)\delta(x')\\
 &= - \left(\frac{ik'}{ik' + \lambda} \right)  \left(\frac{ik}{ik - \lambda}\right)f(0).  
 \end{aligned}
\end{equation}
That is exactly the second contribution \eqref{eq:secondcontdelta} with opposite sign. Thus, the second and the third term cancel out. The fourth contribution ($vv$) to the kernel $2iL(k',k)$ is
\begin{equation*}
\begin{aligned}
 \int dx \int dx' \theta(x') \left( T^{\star}(k')\right)\exp(-ik'x') &\left(2f(x')\frac{\pa \delta(x'-x)}{\pa x'} + \frac{\pa f(x')}{\pa x'}\delta(x'-x) \right) \\ &\times \left( \theta(x) \left( T(k) \right)\exp(ikx)  \right),
 \end{aligned}
 \end{equation*}
which, after similar calculations, results in the fourth and last term
\begin{equation}
i(k + k')  T^{\star}(k') T(k) \int dx' f(x') \theta(x') \exp(ix'(k-k')).    
\end{equation}
 Finally, we can write down the lowest backflow eigenvalue as 

\begin{equation*}
\beta_V(f) = \frac{1}{2\pi}\int_{0}^{\infty}dk \int_{0}^{\infty}dk' \mathcal{\tilde{J^*}}(k')\mathcal{\tilde{J}}(k) L(k',k),
\end{equation*}
with the Hermitian kernel
\begin{eqnarray}\label{kerneldelta} \nonumber
  2L(k',k) &=&  (k + k') \int_{-\infty}^{0} dx' f(x') \exp(ix'(k-k')) \\ \nonumber
  &+&  \frac{\lambda(k'-k)}{(ik - \lambda)}\int_{-\infty}^{0} dx' f(x')\exp(-ix'(k+k')) \\ \nonumber 
  &-& \frac{\lambda(k - k')}{(ik' + \lambda)} \int_{-\infty}^{0} dx' f(x') \exp(ix'(k+k'))\\ \nonumber
  &+&  \frac{\lambda^2 (k+k')}{(ik' + \lambda)(ik - \lambda)}\int_{-\infty}^{0} dx' f(x') \exp(-ix'(k-k')) 
  \\ 
 &-&   \frac{kk'(k + k')}{(ik' + \lambda)(ik - \lambda)}  \int_{0}^{\infty} dx' f(x') \exp(ix'(k-k')).
\end{eqnarray}
Interestingly, the term proportional to $f(0)$, which involves the evaluation of the test function at the defect's location, was cancelled out. That term would be a contribution coming purely from the defect. In the case of the jump-defect, we will see in section~\ref{subsec:backflowinjump} that, if we insist on assigning a value to the wavefunction at the origin, this term is non-zero, due to the discontinuity, and it has a direct connection with the conservation law of the total momentum. In the $\delta$-defect case, as mentioned before in section~\ref{sec:integrable}, there was no additional term we could have added for redefining the total momentum in order to keep it conserved. Coincidentally, what would be a possibly equivalent additional fixing term, coming purely from the defect, for that purpose is zero. The calculation above considered the full time-independent solution $\varphi_k$ including its value at the origin, where the defect was placed, but the jump-defect case will be treated differently.

For the calculation of the lowest backflow eigenvalue $\beta_V(f)$, as the eigenfunction $\mathcal{\tilde{J}}(k)$ is not analytically known, we need to rely upon numerical calculations in order to plot the result. Some graphs for the $\delta$-defect case can be found in section~\ref{sec:results} along with some details of the numerical methods.

\end{subsection}

\begin{subsection}{Backflow in the presence of a jump-defect}\label{subsec:backflowinjump}

Now we consider the backflow calculation for the the jump-defect. However, we have to keep in mind that now our wavefunction $\varphi_k$ has a jump discontinuity at the origin. Specifically, in the $\delta$-defect case, the wavefunction  and its derivative are locally integrable, that is $\varphi_k \in L_{\textrm{loc}}^{1}(\mathbb{R})$ and $\partial_x\varphi_k \in L_{\textrm{loc}}^{1}(\mathbb{R})$. In the jump-defect case, just the wavefunction is locally integrable $\varphi_k \in L_{\textrm{loc}}^{1}(\mathbb{R})$ but not its derivative. Such discontinuities may cause the presence of undefined terms when multiplied by distributions. By avoiding the origin, we avoid this undesirable problem.

Given that now $\varphi_k$ denotes the the solution for the TISE in the presence of a jump-defect. Let us write the full time-dependent jump solution to the Schr\"{o}dinger equation as
\begin{equation}\label{eq:jumptimedep}
\varphi(x,t) = \frac{1}{\sqrt{2}} \int_{-\infty}^{\infty} dk \, \tilde{g}(k) \exp(-iwt)\varphi_{k}(x) 
\end{equation}
with $\tilde{g}$ an arbitrary non-zero smoothly varying function, and the time-independent scattering states in position basis are given by 
 \begin{equation}\label{jumpuvtimeindep}
 \varphi_{k}(x) =
 \left\{
	\begin{array}{ll}
		u_k(x) &=  \exp(ikx), \quad x<0 \\[0.5cm]
		v_k(x) &= \left(  \dfrac{k+i\alpha}{k - i\alpha}\right)\exp(ikx), \quad x>0 ,
	\end{array}
\right.
\end{equation}
where the reflection coefficient $R(k)=0$ and the transmission coefficient $T(k)$ for the jump-defect is explicit. While the jump-defect connects the theory on the left of the origin ($x=0$) with the theory on the right, we do not assign a definite value for the wavefunction at the defect's position. Because of that, we do not have all the corresponding four contributions calculated in section~\ref{subsect:deltadefect}, but only two of them, $uu$ and $vv$. In fact, if we do a similar calculation to that of section~\ref{subsec:conservationlaws}, we only split our integration \eqref{eq:innerintegral} in a left part ($-\infty < x <0 $) and a right part ($0<x< \infty$), corresponding to contributions purely from  $u$ and purely from $v$, respectively, and we avoid crossing the discontinuity. 

We concentrate our attention to the time-independent part $\varphi_k(x)$ composed of \eqref{jumpuvtimeindep} and, for that, the asymptotic backflow constant of the jump-defect is
\begin{equation}
\begin{aligned}
\beta_V(f) =  \frac{1}{2\pi}\int_{0}^{\infty}dk \int_{0}^{\infty}dk' \mathcal{\tilde{J^*}}(k')  \mathcal{\tilde{J}}(k)  \int dx  \int dx' &  \left[  \theta(-x')u^{\star}_{k'}(x')  J(f)(x',x)\theta(-x)u_{k}(x)  \right. \\ & \left.
\phantom{m}+ \theta(x')v^{\star}_{k'}(x')J(f)(x',x)\theta(x)v_{k}(x) \right]. 
\rule[-1.0em]{0pt}{0pt}
\end{aligned} 
\end{equation}
 Thus we have two contributions where each term is expanded by the insertion of the $J(f)(x',x)$ and simplified after integration.
Let us focus only on the spatial integrals, namely the kernel $2iL(k',k)$. The first contribution ($uu$) is
\begin{equation*}
 \int dx \int dx' \theta(-x')u^{\star}_{k'}(x')\left(2f(x')\frac{\pa \delta(x'-x)}{\pa x'} + \frac{\pa f(x')}{\pa x'}\delta(x'-x) \right) \theta(-x)u_k(x),
 \end{equation*}
which is integrated by parts to become
 \begin{equation*}
 \begin{aligned}
      \int dx \int dx' \left\{-\frac{\pa}{\pa x'}\right. & \left. \left(\theta(-x') u^{\star}_{k'}(x')2f(x') \right) \delta(x'-x)\theta(-x)u_k(x) \right. \\ &+ \left. \theta(-x')u^{\star}_{k'}(x')\frac{\pa f(x')}{\pa x'} \delta(x'-x) \theta(-x) u_k(x) \right\}.
     \end{aligned}
 \end{equation*}
After one integration is carried out, it gives
 \begin{equation*}
 \begin{aligned}
 \int dx' &\left\{ \theta(-x')u^{\star}_{k'}(x')2f(x') u_k(x')\left[-\delta(x') + \theta(-x')ik \right]  - \theta(-x')(-ik')u^{\star}_{k'}(x')f(x')\theta(-x')u_k(x') \right. \\ & \left.  \phantom{m} + \left(\delta(x')u^{\star}_{k'}(x')f(x')\theta(-x')u_k(x') \right)  - \theta(-x')u^{\star}_{k'}(x')f(x')u_k(x') \left(-\delta(x') + ik \theta(-x')\right)
\right\},
\end{aligned}
 \end{equation*}
where we have used $\pa \theta(-x')/ \pa x' = - \delta (x')$.
Hence, the first contribution term is
\begin{equation}
   i(k + k') \int_{-\infty}^0 dx' \exp(ix'(k-k'))f(x').  
\end{equation}
The second contribution ($vv$) to $2iL(k',k)$ will involve similar calculations. It is given by

\begin{equation*}
 \int dx \int dx' \theta(x')v^{\star}_{k'}(x')\left(2f(x')\frac{\pa \delta(x'-x)}{\pa x'} + \frac{\pa f(x')}{\pa x'}\delta(x'-x) \right) \theta(x)v_k(x),
 \end{equation*}
 which, on integrating by parts, yields
 \begin{equation*}
 \begin{aligned}
    \int dx \int dx' & \left\{ -\frac{\pa}{\pa x'} \left(\theta(x')T^{\star}(k') \exp(-ik'x')2f(x') \right) \delta(x'-x)\theta(x)T(k)\exp(ikx) \right. \\
    & \left. \phantom{mm}  +\theta(x')T^{\star}(k')\exp(-ik'x')\frac{\pa f(x')}{\pa x'} \delta(x'-x) \theta(x)T(k) \exp(ikx)\right\}. \end{aligned} 
     \end{equation*}
After one integration is carried out, it becomes
 \begin{equation}
 i(k+k') \int_{0}^{\infty} dx' f(x')  T^{\star}(k') T(k)\exp(ix'(k-k')).   
 \end{equation}
Finally, we can write the lowest backflow eigenvalue of the operator $J_V(f)$ as 
\begin{equation*}
\beta_V(f) = \frac{1}{2\pi}\int_{0}^{\infty}dk \int_{0}^{\infty}dk' \mathcal{\tilde{J^*}}(k')\mathcal{\tilde{J}}(k) L(k',k),
\end{equation*}
with the kernel 
\begin{equation} \label{kerneljump} 
\begin{aligned}
 2L(k',k) = (k + k') & \int_{-\infty}^{0} dx' f(x') \exp(ix'(k-k'))  \\
 & \phantom{m}+ \frac{\left(kk' + i\alpha(k'-k) + \alpha^2 \right)}{(k'+i\alpha)(k-i\alpha)}(k+k') \int_{0}^{\infty} dx' f(x') \exp(ix'(k-k')),   
 \end{aligned} 
\end{equation}
which is a Hermitian kernel, and $\mathcal{\tilde{J}}(k)$ is the eigenfunction, in momentum space, associated with the lowest eigenvalue of the integral operator $J_{V}(f)$. This expression \eqref{kerneljump} was worked out for the non-conserved situation where we have not introduced any fixing term to conserve the probability current. In physical situations, we are interested in conserved quantities, and our jump-defect was specially devised for allowing conservation laws.

In section~\ref{sec:integrable}, we have established the condition for having a conserved total momentum $P_c$ associated with a particular momentum density. The probability current is intimately related to the total momentum since
\begin{equation}
    \int j_{\psi}(x) dx = \braket{\hat{P}},
\end{equation}
where we have set $\hbar = m =1$, and $\hat{P}$ is the momentum operator. We can, therefore, interchangeably, refer to either momentum density or, equivalently, probability current density. 
In particular, Eq. \eqref{Termconservemomentum} determines the
adjusting term for obtaining a conserved probability current. The
adjustment needs to be written in terms of a kernel, in momentum space, such that it can be added to the kernel $L(k',k)$ in \eqref{eq:innerintegral}. From \eqref{eq:jumptimedep}, we can write
the time-independent solution at the left of the defect as
\begin{equation}
\begin{aligned}
    u(x) &= \frac{1}{\sqrt{2\pi}}\int dk \tilde{g}(k) u_k(x), \\
    u^{\star}(x) &= \frac{1}{\sqrt{2\pi}}\int dk' \tilde{g}^{\star}(k') u^{\star}_{k'}(x),
    \end{aligned}
\end{equation}
and similarly for the solution $v$ at the right of the deft. Hence, after introducing the required projectors $E_+$ for right-movers, the adjustment expression is
\begin{equation}
 \left. \frac{i}{2}E_+(u^{\star}v - v^{\star}u)E_+\right|_{x=0} =  \frac{i}{4\pi} \int_{0}^{\infty} \int_{0}^{\infty} dk' dk \, \tilde{g}^{\star}(k')\tilde{g}(k) \left( \frac{2i\alpha(k+k')}{(k-i\alpha)(k'+i\alpha) }  \right).
\end{equation}
Note, section~\ref{sec:integrable} has no reference to the smearing
process with a positive test function $f$ for producing spatial averaged quantities as introduced in our discussion of the quantum backflow in section~\ref{sec:qbackflow}. With our test function being a function only of the position, rather than time, the corresponding spatial averaged quantity is exactly what needs to be added to the expectation value $\expval{J_V(f)}_{\psi}$, from \eqref{eq:interactingcurrent}, to give us the corresponding expectation value denoted by
\begin{equation}\label{eq:conservedexpectvalue}
   \expval{J_c{_V(f)}}_{\psi} = \expval{J_V(f)}_{\psi} + \frac{i}{4\pi} \int_{0}^{\infty} \int_{0}^{\infty} dk' dk \, \tilde{\psi}^{\star}(k')\tilde{\psi}(k) \left( \frac{2i\alpha(k+k')}{(k-i\alpha)(k'+i\alpha) }  \right)f(0),
\end{equation}
which is now actually related to the conserved probability current. Here the term
$\expval{J_V(f)}_{\psi}$ only includes the kernel's contributions \eqref{kerneljump} which do not come from the defect. The defect's contribution is only taken into consideration when we impose the conservation of the probability current associated with the physical system we want to describe, which does not include the defect a priori. Finally, the expression \eqref{eq:conservedexpectvalue} can be written as
\begin{equation}
\begin{aligned}
  \expval{J_c{_V(f)}}_{\psi} = \frac{1}{2\pi} \int_{0}^{\infty} \int_{0}^{\infty} & dk' dk \, \tilde{\psi}^{\star}(k')\tilde{\psi}(k) \left[ (k + k') \int_{-\infty}^{0} dx' f(x') \exp(ix'(k-k')) 
 \right.  \\
 & \left. + \frac{\left(kk' + i\alpha(k'-k) + \alpha^2 \right)}{(k'+i\alpha)(k-i\alpha)}(k+k') \int_{0}^{\infty} dx' f(x') \exp(ix'(k-k'))  \right.  \\ & \left.
  \phantom{mm}-\left( \frac{\alpha(k+k')}{(k-i\alpha)(k'+i\alpha) }  \right)f(0)
                \right].
 \end{aligned}
\end{equation}
From this one only needs to take the infimum over the functions $\psi$, as in \eqref{eq:beta0}, in order to obtain the lowest backflow eigenvalue $\beta_V(f) = \inf \expval{J_c{_V(f)}}_{\psi}$ of the probability current operator $J_c{_V(f)}$ in the presence of the jump-defect. Once we have simplified the kernel, we again need to rely upon the numerical calculations as the eigenfunction $\mathcal{\tilde{J}}(k)$ is not analytically known.

As we mentioned, non-removable discontinuities pose difficulties to the products with Dirac measure $\delta$, which is a Radon measure rather than a locally integrable function. In particular, a distributional product such as
$$ \left< \varphi\delta,f \right> = \left< \delta,\varphi f \right> = \varphi(0)f(0),$$
with a function $\varphi$  discontinuous at the origin and test function $f$, is undefined. If we had insisted on following the same calculations as done in the $\delta$-defect case, crossing the origin as a range of integration, and assigning a particular value for the Heaviside function at the origin, such as $\theta(0)= 1/2$, in order to have a meaningful way to interpret the difficulties caused by the discontinuity, we would get an extra non-zero term, corresponding to the combination of $uv$ and $vu$ contributions,
\begin{equation}\label{extra-term}
\frac{2\alpha(k+k')}{(k'+i\alpha)(k-i\alpha)}f(0)
\end{equation}
to the kernel \eqref{kerneljump}. In the $\delta$-defect case with a continuous wavefunction, the corresponding contribution is zero as previously discussed. This term \eqref{extra-term} would come after a delta $\delta(x')$ is integrated out and it has, therefore, support at the origin. In the Lagrangian \eqref{lagrangian}, we have made it clear that what is defined only at the defect's position is a contribution purely from the defect itself rather than the theory of the left or the theory of the right semi-infinite lines. Incidentally, it is curious that this term \eqref{extra-term}, apart from a sign, is exactly the fixing term in \eqref{eq:conservedexpectvalue} that we need for the conserved probability current.

\end{subsection}

\begin{sidewaysfigure}[!htb]
    \includegraphics[width=1.0\textwidth]{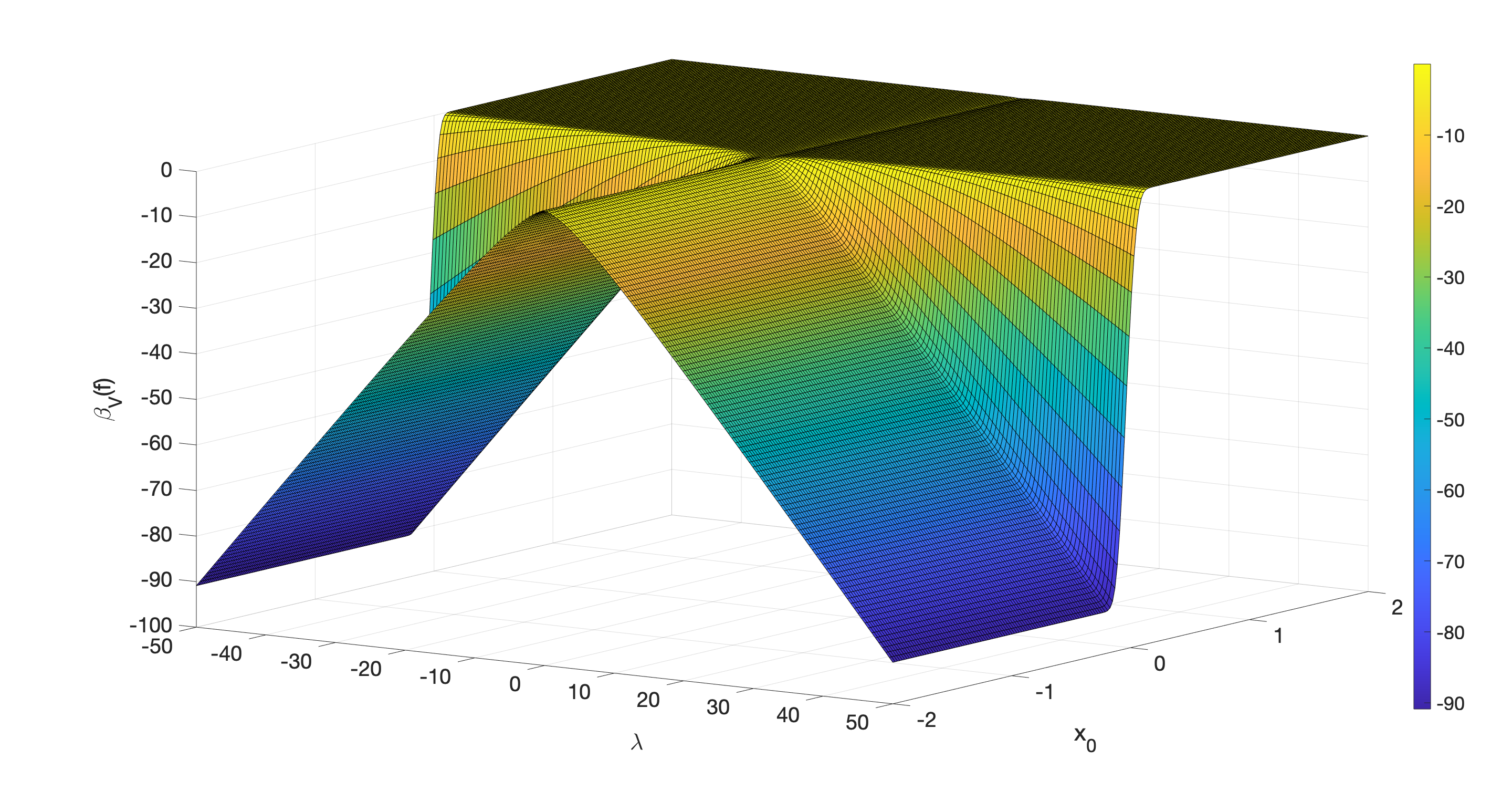}
    \caption{Probability current lowest eigenvalue for $\delta$-defect, $P_{\textrm{cutoff}}=200$, $N=2000$.}
    \label{fig:landscapedelta}
\end{sidewaysfigure}

\begin{sidewaysfigure}[ht]
    \includegraphics[width=1.0\textwidth]{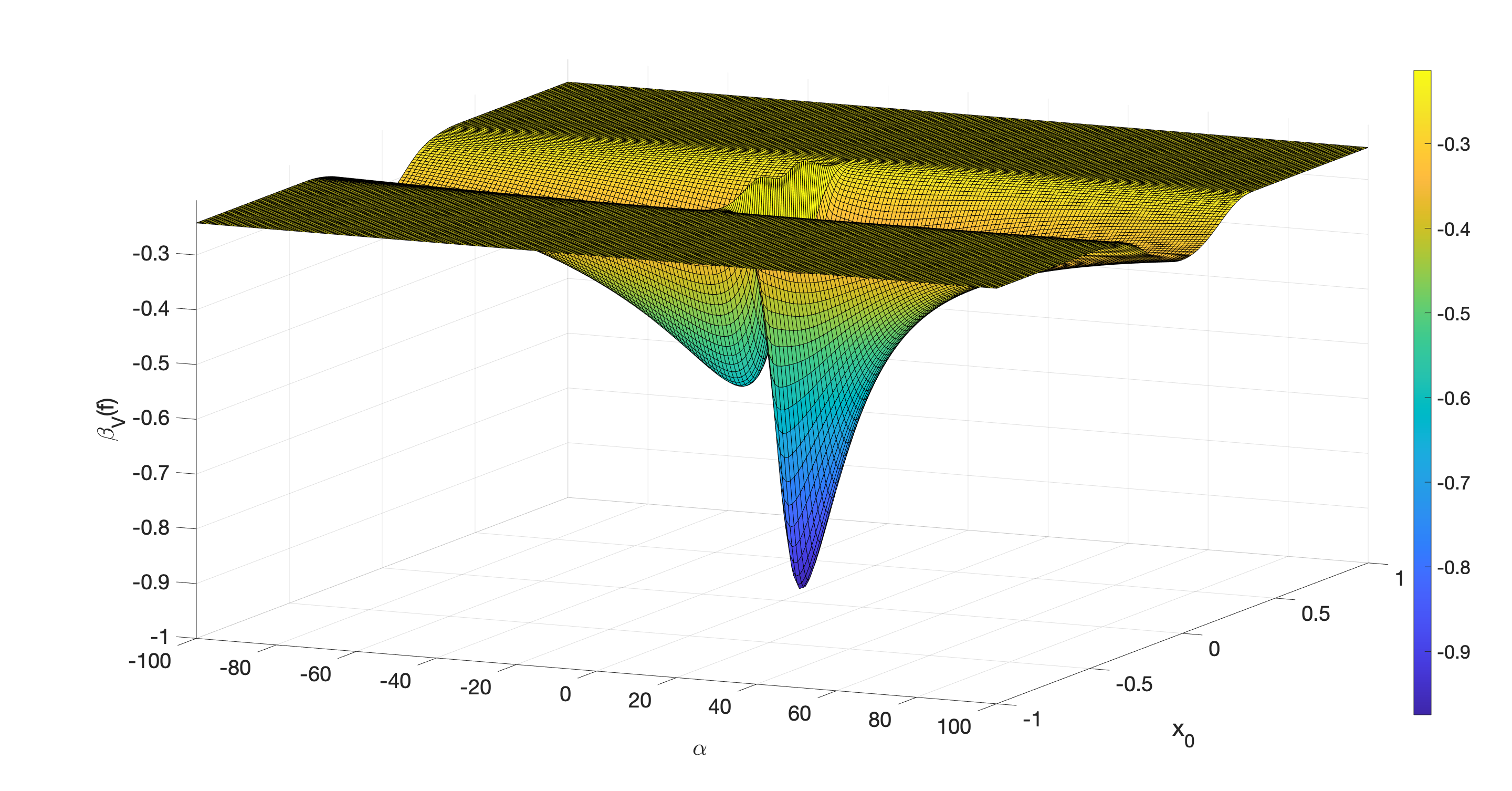}
    \caption{Probability current lowest eigenvalue, $P_{\textrm{cutoff}}=200$, $N=2000$.}
    \label{fig:landscapenonconserved1}
\end{sidewaysfigure}

\begin{section}{Results}\label{sec:results}

\begin{subsection}{Numerical calculations}

We have adapted the basic numerical methods of \cite{Henning} (where one can find the essential numerical description with a Java program) for a FORTRAN 90 program with some changes in regards to the method of integration and the calculation of the lowest eigenvalue of a complex Hermitian matrix $M$. The discontinuities due to the defects are also taken into account in the numerical integration. For that, the libraries used were QUADPACK \cite{quadpack} and EISPACK \cite{eispack}, respectively. Here is a summary of the meaning of each relevant variable to understand the plots presented in this work.

For the numerical calculations, the discretization of an infinite-dimensional operator $T$ on $L^{2}(\mathbb{R}_+, dk)$ with kernel $K$ given by
\begin{equation}
 K(k',k) =   \frac{1}{2\pi}L(k',k)  = \frac{i}{4\pi} \int dx f(x) \left(\pa_x \varphi^{\star}_{k'}(x)\varphi_k(x) - \varphi^{\star}_{k'}(x)\pa_x \varphi_k(x) \right) 
\end{equation}
into a N$\times$N-matrix $M$ is characterized by the parameter N, the number of equally spaced steps which divide the momentum interval $[0,P_{\textrm{cutoff}}]$, where the upper-limit cutoff of the integrations \eqref{interactingexpectationvalue} in $k$ and $k'$ is denoted by $P_{\textrm{cutoff}}$. The components of such a matrix can be written as
\begin{equation}
    M_{ij} = \hscalar{\psi_i}{T\psi_j} = \int dk' \int dk \, \tilde{\psi}_i(k')K(k',k)\tilde{\psi}_j(k) \approx \frac{P_{\textrm{cutoff}}}{N}K(k_i,k_j),
\end{equation}
where $\tilde{\psi}_i \, (i \in \mathbb{N} \, \, \vert \, i = (0, \ldots, N-1))$ are orthonormal step functions supported on the corresponding interval, and mid-points $k_i = \left(i + 1/2\right)\left(P_{\textrm{cutoff}}/N\right)$. The adoption of a cutoff $P_{\textrm{cutoff}}$ is consistent with the fact that the lowest backflow eigenvector decays at large momentum. 

The positive test function chosen for the spatial average of the probability current was a Gaussian 
\begin{equation}\label{testfunction}
  f(x) = \frac{1}{\sigma\sqrt{2\pi}} \exp\left( -\frac{(x-x_0)^2}{2\sigma^2} \right) , 
  \end{equation}
with width $\sigma=0.1$ centered at the position $x_0$ of measurement where a spatially extended detector is located and supported on the interval $ x \in [x_0-8\sigma, x_0+8\sigma]$ . Therefore, for each $x_0$ we have a matrix $M$ for which the lowest eigenvalue needs to be calculated. We have restricted our position of measurement to $ x_0 \in [-2,2]$ in the case of the delta-defect and to $x_0 \in [-1,1]$ in the jump-defect case because, as we move away from the jump-defect's location, the lowest eigenvalue approaches the free case value $\beta_0(f) \approx -0.241$ for this choice of test function.

Although essentially the same, the numerical analysis done for the conserved probability current involves an extra step, which is the addition of a fixing term to the non-conserved one such that the fixing term allows the conservation law to hold. Specifically, the fixing term in the presence of a jump-defect,

\begin{equation}
    -\frac{1}{2\pi} \frac{\alpha(k+k')}{(k-i\alpha)(k' +i\alpha)}f(0),
\end{equation}
is added to $K(k',k)$ to compose a new kernel denoted by $K_c(k',k)$, which is associated with a conserved quantity. The discretization process now involves that new kernel, and the FORTRAN program is asked to calculate the lowest eigenvalue $\beta_V(f)$ of the corresponding N$\times$N-matrix $M$.

\end{subsection}

\subsection{$\delta$-defect case}

The backflow calculation in the presence of a $\delta$-defect was analysed in section~\ref{subsect:deltadefect}, and the corresponding kernel was analytically simplified to expression \eqref{kerneldelta}. Here, we present additional numerical results to those reported in \cite{Henning}, where the defect parameter was restricted to $|\lambda|=1$. All the graphs refer to the probability current operator smeared with a Gaussian test function. Specifically, the graphs show the lowest backflow eigenvalue $\beta_V(f)$ against the center $x_0$ of the averaging Gaussian function $f$. See the following figure~\ref{fig:deltaposition1} and figure~\ref{fig:deltaposition2} where we vary the parameter $\lambda$ for displaying its behaviour under the strengthening or weakening of the interaction. In particular, in the limit $\lambda \rightarrow \pm \infty$, it becomes a purely reflecting situation, equivalent to a boundary theory. Naturally, when $\lambda \rightarrow 0$ the  interaction-free case is obtained. For the free case, the lowest eigenvalue is represented by the line $\beta_0(f) \approx -0.241$. As shown by figure~\ref{fig:deltaposition1}, there is a maximum of the lowest eigenvalue, in the attractive case, close to the defect's location when $|\lambda| < 1$. Moreover, the maximum seems to peak when the defect parameter is $\lambda=-1/2$. Despite not being included in this report, a few of other parameters around $\lambda=-1/2$ ($\lambda = -0.40, -0.45,-0.55,-0.60$) were explored and suggested that the maximum indeed peaks at $\lambda =-1/2$. This is a new observation. We do not have a physical explanation for it, but it is worth exploring in the future. Increasing its absolute value ($|\lambda|>1$) causes the attractive and repulsive cases to approach each other.

Additionally to the two-dimensional plots, we have varied the parameter $\lambda$ over a wide range for displaying a three-dimensional picture, figure~\ref{fig:landscapedelta}, to illustrate how the lowest backflow eigenvalue is affected in the presence of the $\delta$-defect. This can be compared to the situation when the jump-defect parameter $\alpha$ varies, figures~\ref{fig:landscapenonconserved1}, \ref{fig:landscapenonconserved2}, \ref{fig:landscapeconserved1} and \ref{fig:landscapeconserved2} in the Appendix.

\newpage

\begin{figure}[!htb]\centering
 \begin{subfigure}{\linewidth}
\includegraphics[width=0.95\linewidth]{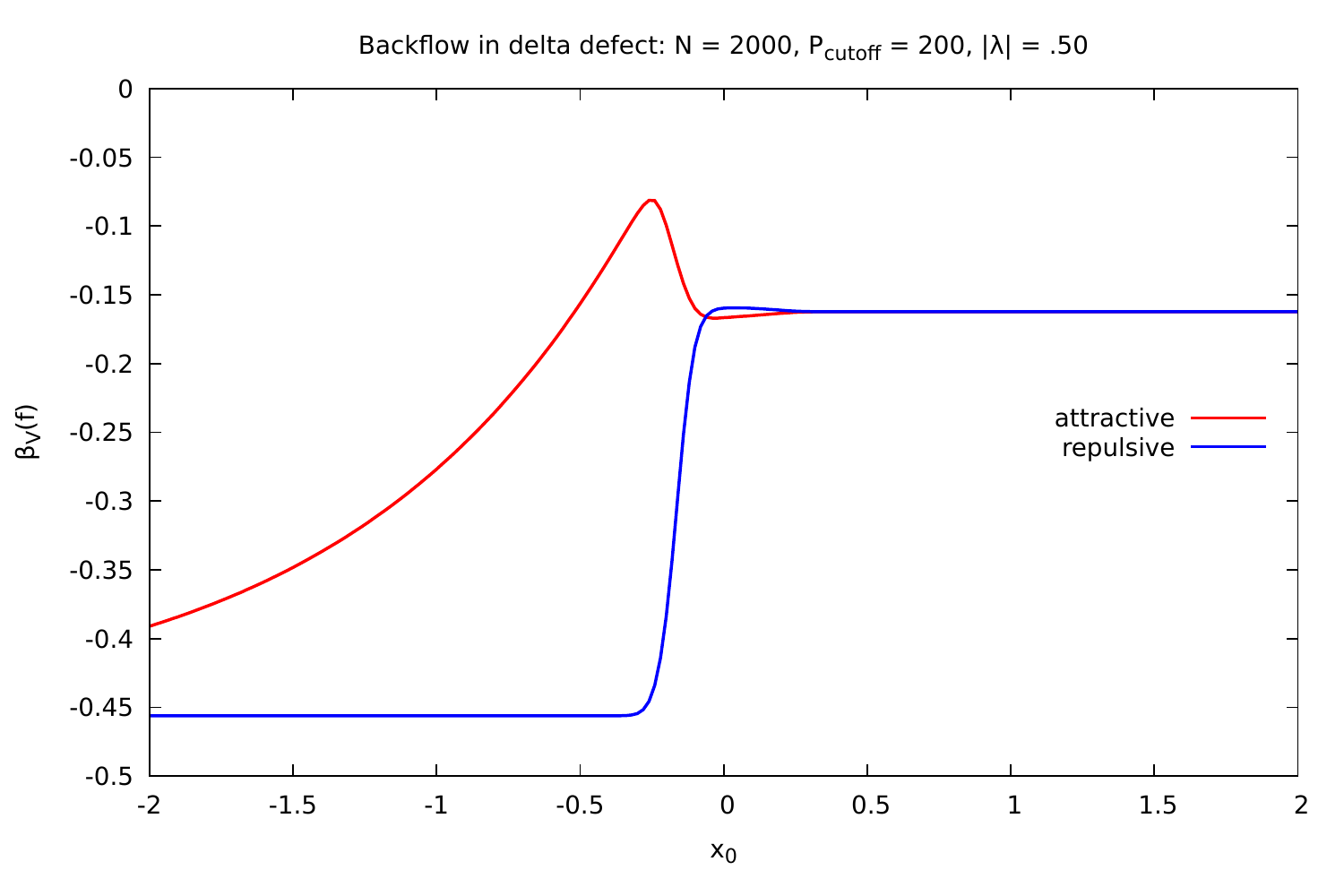} \caption{}
\end{subfigure} \hfill

\begin{subfigure}{\linewidth}
\includegraphics[width=0.95\linewidth]{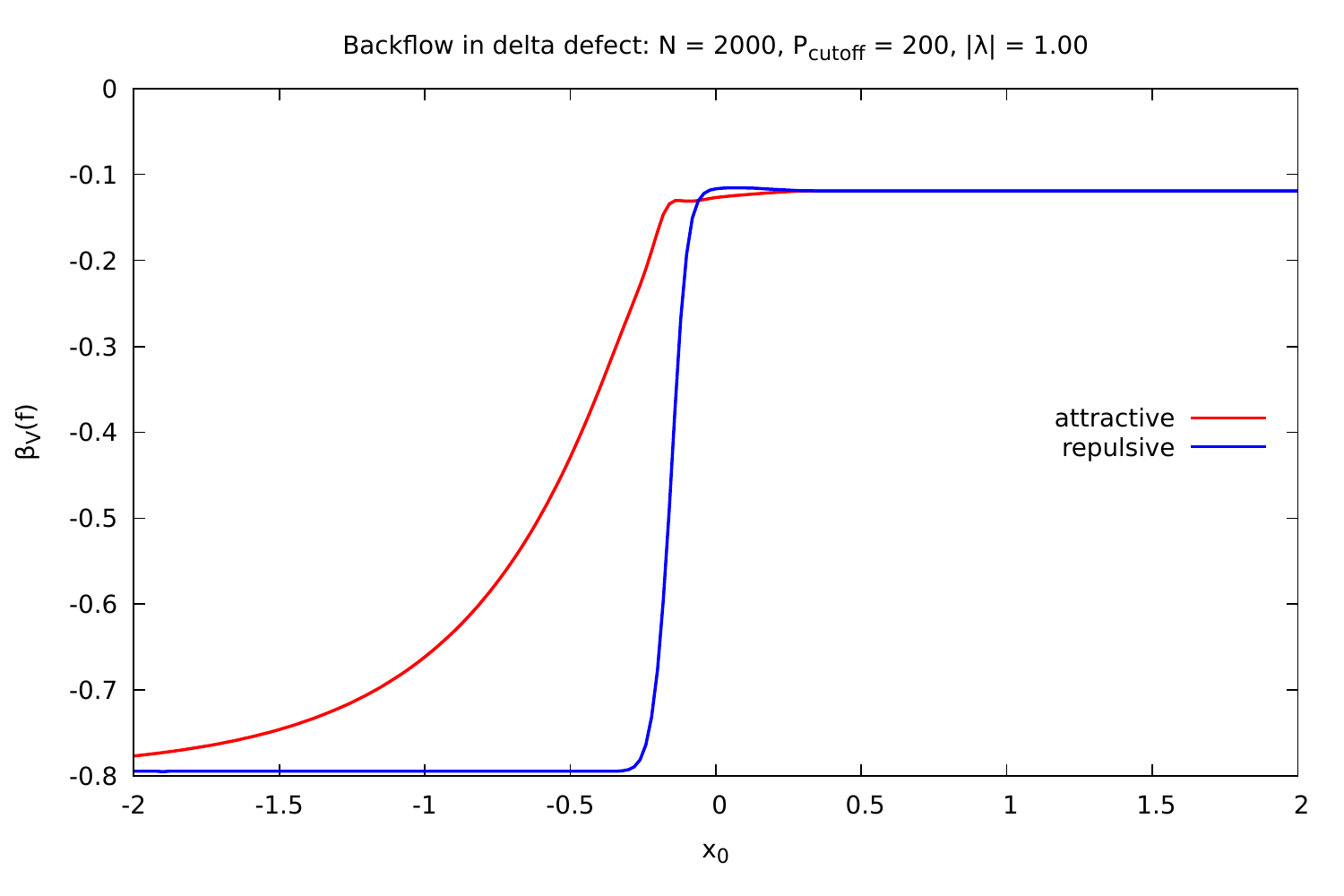}\caption{} 
\end{subfigure}\hfill 
 \caption{Lowest backflow eigenvalue of the current operator. For which (a) $|\lambda| =0.5$. \\ (b) $|\lambda| =1.0$. } \label{fig:deltaposition1}
  \end{figure}

\begin{figure}[!htb]\centering
 \begin{subfigure}{\linewidth}
\includegraphics[width=0.95\linewidth]{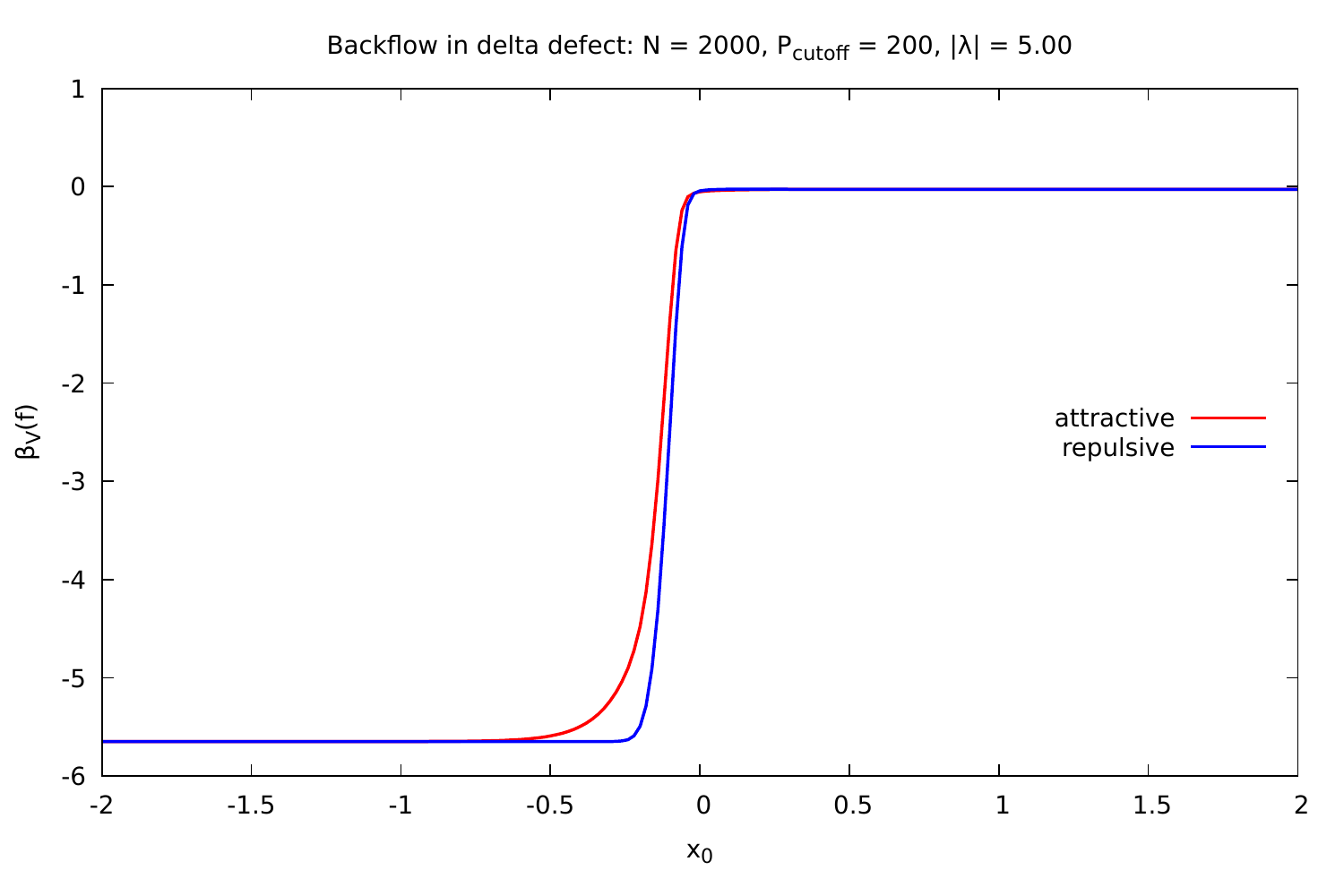}\caption{}
\end{subfigure} \hfill

\begin{subfigure}{\linewidth}
\includegraphics[width=0.95\linewidth]{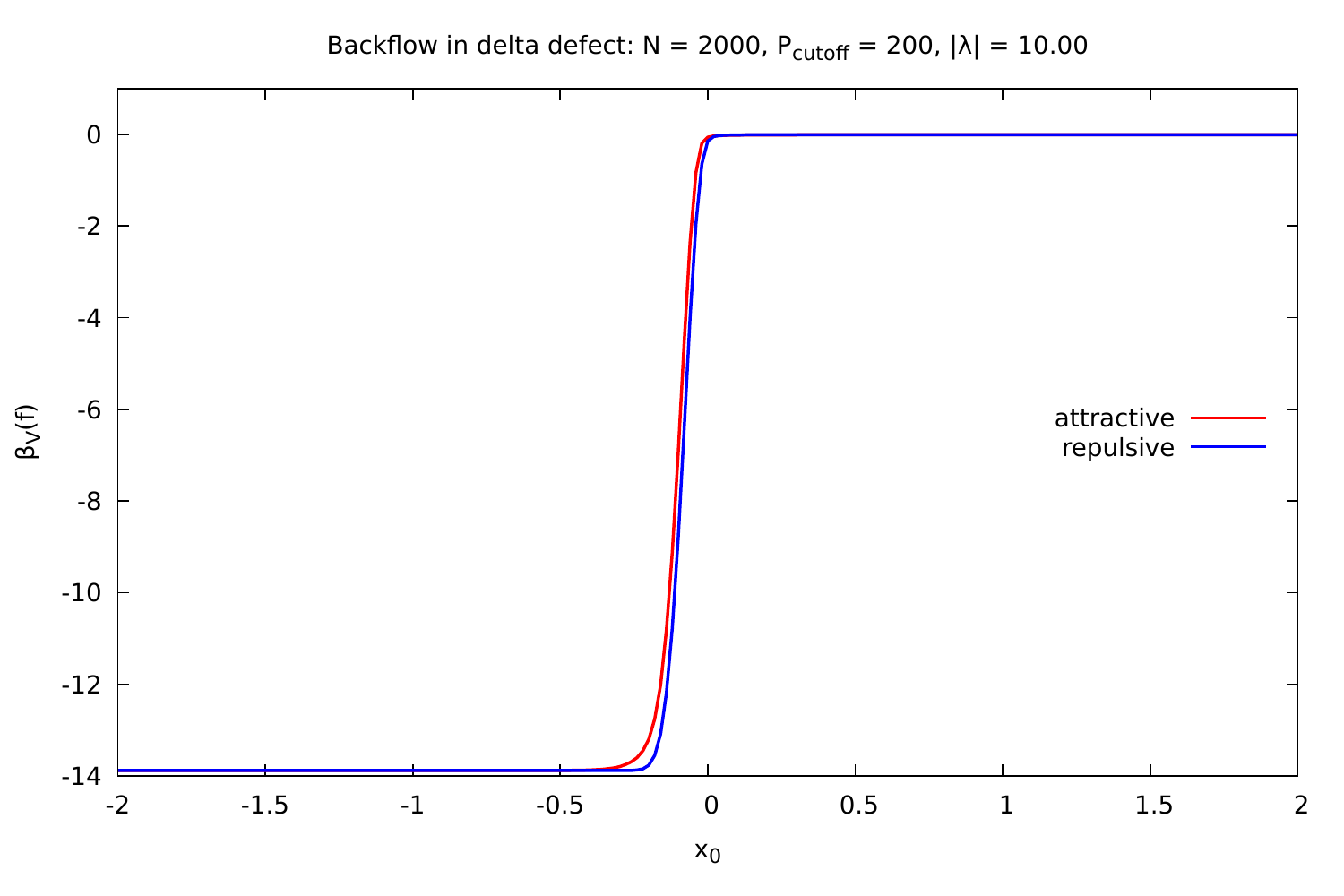} \caption{} 
\end{subfigure} 
 \caption{Lowest backflow eigenvalue of the current operator. For which (a) $|\lambda| =5.0$. \\ (b) $|\lambda| =10.0$.} \label{fig:deltaposition2}
  \end{figure} 

\clearpage
\newpage

\begin{subsection}{Jump-defect case}

For a purely-transmitting defect, the solution $\varphi_k$ with an asymptotic incoming right-mover maintains itself as a right-mover also after scattering off the defect. This is not the case for the $\delta$-defect that has a mixture of right-movers and left-movers as a result of being scattered by the defect. In this sense, the reflectionless P\"{o}schl-Teller potential is more similar to the jump-defect than the $\delta$. However, for the P\"{o}schl-Teller potential, the backflow effect is smaller inside the interaction region than in the free case \cite{Henning}. That this is not true in the jump-case can be seen from the figures in this section. In fact, at the defect's location, the effect can be either smaller or bigger depending on the magnitude of the parameter $\alpha$.

Several graphs of the backflow lowest eigenvalue in the presence of the jump-defect were plotted below. All the graphs refer to the probability current operator smeared with a Gaussian test function $f$. Specifically, as mentioned in the $\delta$-defect case, the graphs show the lowest eigenvalue against the position of measurement $x_0$ where $f$ is centered. Our main freedom to be tuned is the parameter $\alpha$ corresponding to the strength of the defect. Unlike the Dirac $\delta$-defect, or other explicit potential functions, the jump-defect has a parameter that can not be clearly distinguished as attractive or repulsive according to its sign, being either positive or negative, respectively. As particular cases, $\alpha =0$ gives the expected free case represented by a constant horizontal line $\beta_0(f) \approx -0.241$ and the limiting cases $\alpha \rightarrow \pm \infty $ also approach the free backflow eigenvalue $\beta_V(f) \rightarrow \beta_0(f)$. We already expected this as the solutions $\varphi_k$ for the limiting cases $\alpha \rightarrow \pm \infty $ are related to the free case by only a global phase, but the probability current density has products of the solution wavefunction with its complex conjugated spatial derivative. Thus, in the limit, their lowest backflow eigenvalue is the same as the free case.

Initially, for small absolute values of the parameter $\alpha$, the lowest backflow eigenvalues has some symmetry between the positive and negative parameter values, figure~\ref{fig:alpha001and02}. Slightly increasing $\vert \alpha \vert$, $\beta_V(f)$ of the associated conserved probability current starts to show a distinctly different behaviour between the positive and the negative values of $\alpha$, see figure~\ref{fig:alpha1and4}. As its absolute value increases, the graphs become more similar in terms of the magnitudes of the lowest backflow eigenvalue. However, as indicated by the plots, both  positive $\alpha > 0$ and negative $\alpha < 0$ seem to unveil some stationary points, and, in some cases, while a positive parameter shows three of these points, the corresponding negative parameter can show up to five stationary points, figure~\ref{fig:alpha9and10}. With successive increases of the parameter's absolute value $|\alpha|$, the graphs tend to become more similar again. In particular, both positive and negative values show the same number of stationary points, though when one has a minimum the other one has a maximum and vice-versa, figure~\ref{fig:alpha20and50}. Whilst the non-conserved current develops a persistent trough for both positive and negative parameters, the conserved one develops a mixture of troughs and bumps as shown by Figs.~\ref{fig:alpha1and4}, \ref{fig:alpha9and10} and \ref{fig:alpha20and50}. As numerical results for the rectangular potential in \cite{Henning} suggest, bound states might contribute towards these bumps. It is worth mentioning that, from \eqref{travellingwave}, it is possible to see the existence of bound states associated with the jump-defect for either $k=i\alpha$ or $k=-i\alpha$. The respective bound states can then be described by the following solutions
\begin{equation}
u=0, \, v=v_0\exp(i\alpha^2t/2 -\alpha x), \, (k=i\alpha); \quad u=u_0\exp(i\alpha^2t/2 +\alpha x), \, v=0, \, (k=-i\alpha),    
\end{equation}
which are clearly square-integrable solutions (provided $\alpha > 0$) \cite{Ed2006}.

\begin{figure}[!htb]\centering
  \begin{subfigure}{\linewidth}
  \includegraphics[width=0.95\linewidth]{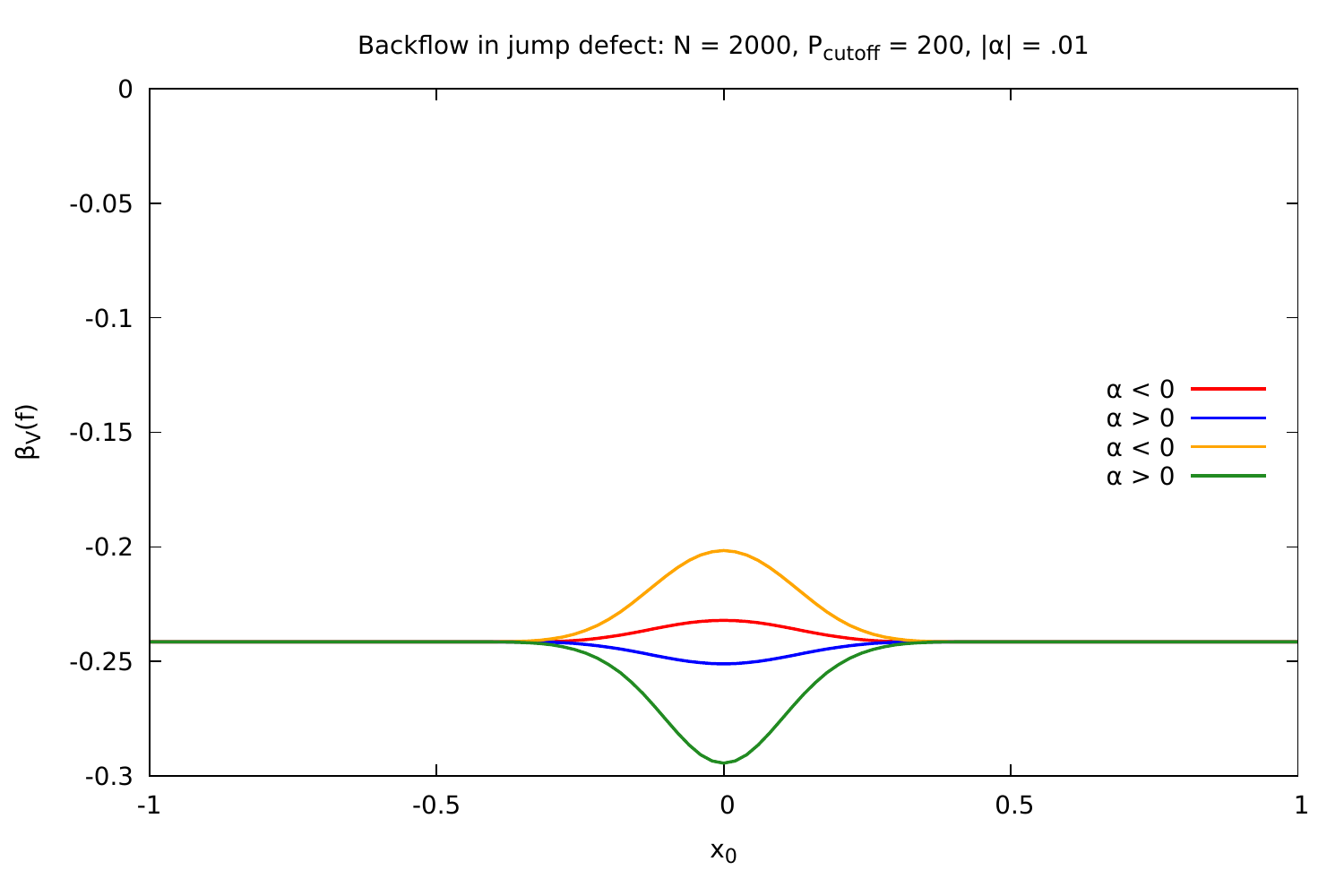}\caption{}
\end{subfigure} 
\begin{subfigure}{\linewidth}
  \includegraphics[width=0.95\linewidth]{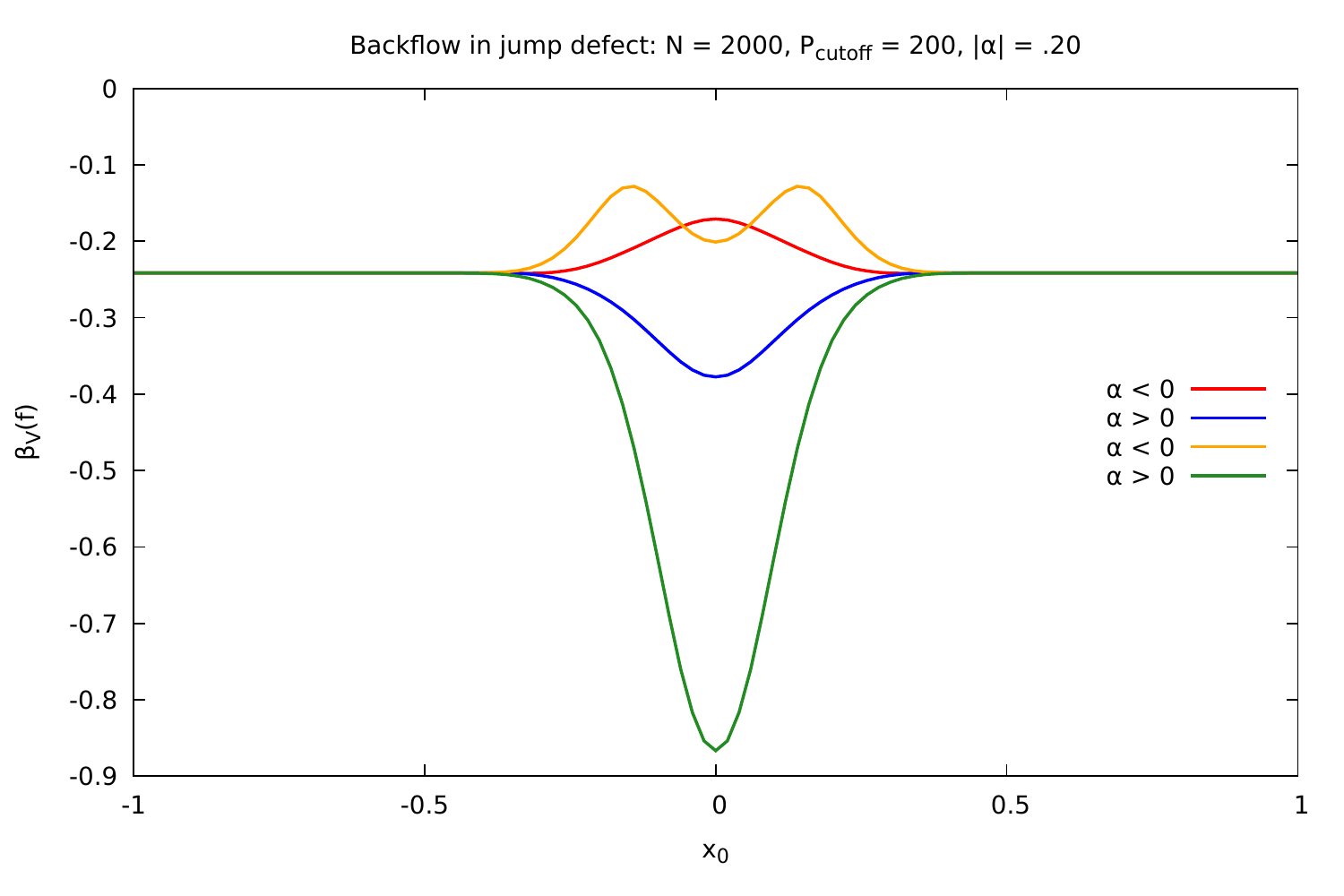}
\caption{}
 \end{subfigure}
   \caption{Lowest backflow eigenvalue of the current operator. \!\textcolor{red}{\textbf{Red}}/\textcolor{blue}{\textbf{blue}} refer to the non-conserved probability current. \textcolor{orange}{\textbf{Yellow}}/\textcolor{OliveGreen}{\textbf{green}} refer to the conserved one. (a) $|\alpha|  = .10$. (b) $|\alpha|  = .20$. } \label{fig:alpha001and02}
  \end{figure}

\begin{figure}[!htb]\centering
  \begin{subfigure}{\linewidth}
  \includegraphics[width=0.95\linewidth]{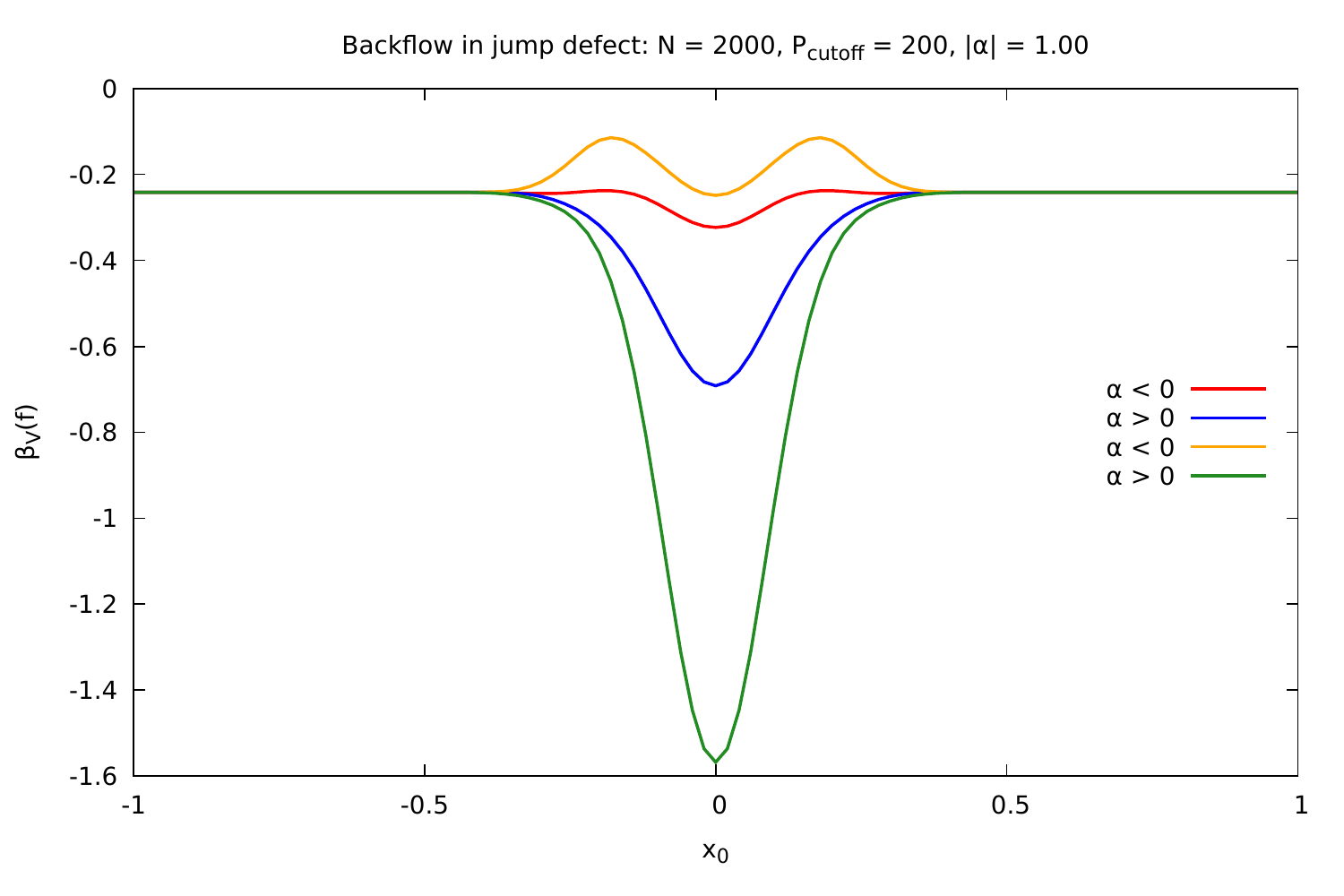} \caption{}
 \end{subfigure} 
  \begin{subfigure}{\linewidth}
  \includegraphics[width=0.95\linewidth]{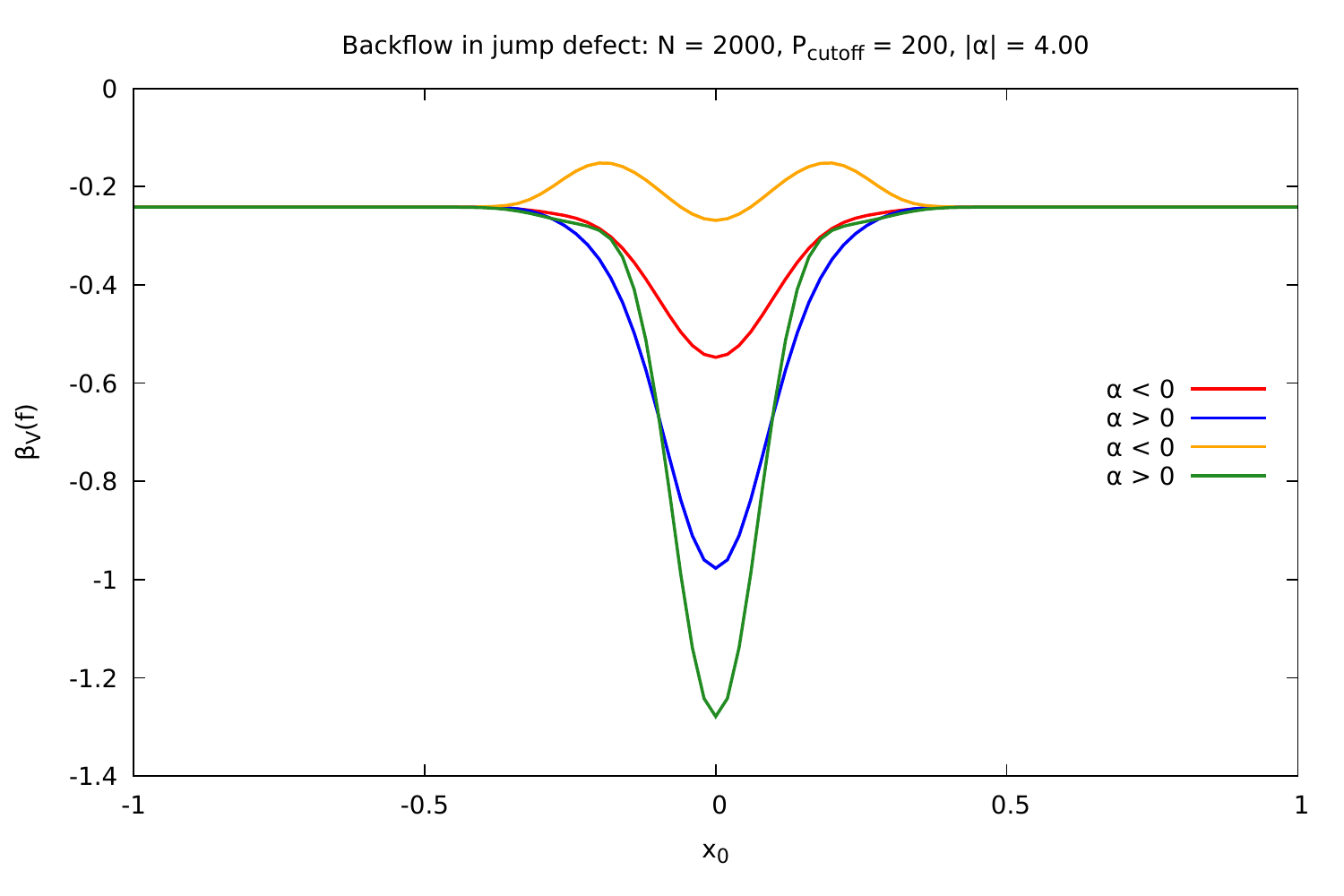} 
\caption{}
  \end{subfigure} 
   \caption{Lowest backflow eigenvalue of the current operator. \!\textcolor{red}{\textbf{Red}}/\textcolor{blue}{\textbf{blue}} refer to the non-conserved probability current. \textcolor{orange}{\textbf{Yellow}}/\textcolor{OliveGreen}{\textbf{green}} refer to the conserved one. (a) $|\alpha|  = 1.0$. (b) $|\alpha|  = 4.0$.} \label{fig:alpha1and4}
  \end{figure}

\begin{figure}[!htb]
  \begin{subfigure}{\linewidth}
  \includegraphics[width=0.95\linewidth]{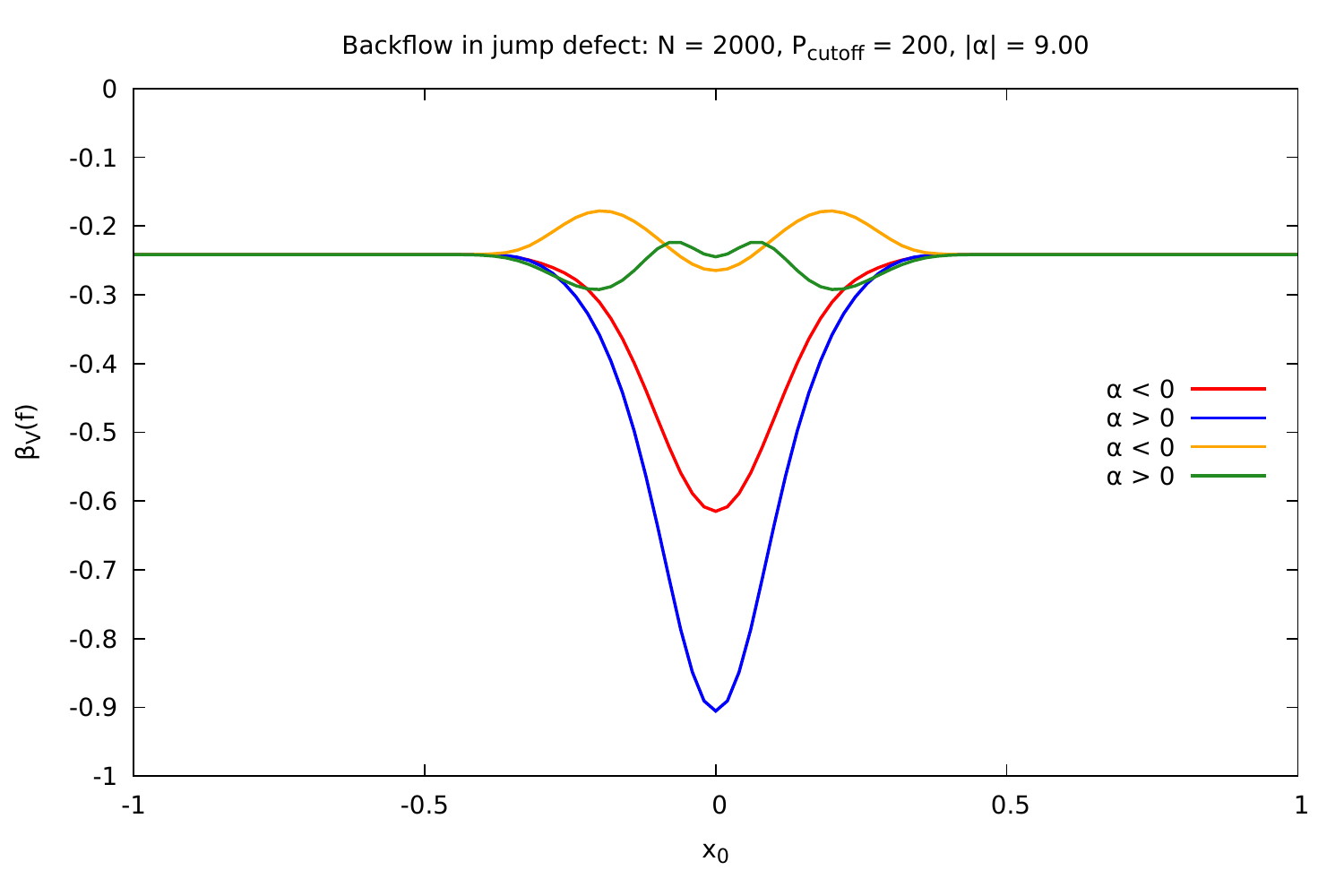}\caption{}
  \end{subfigure}
  \begin{subfigure}{\linewidth}
  \includegraphics[width=0.95\linewidth]{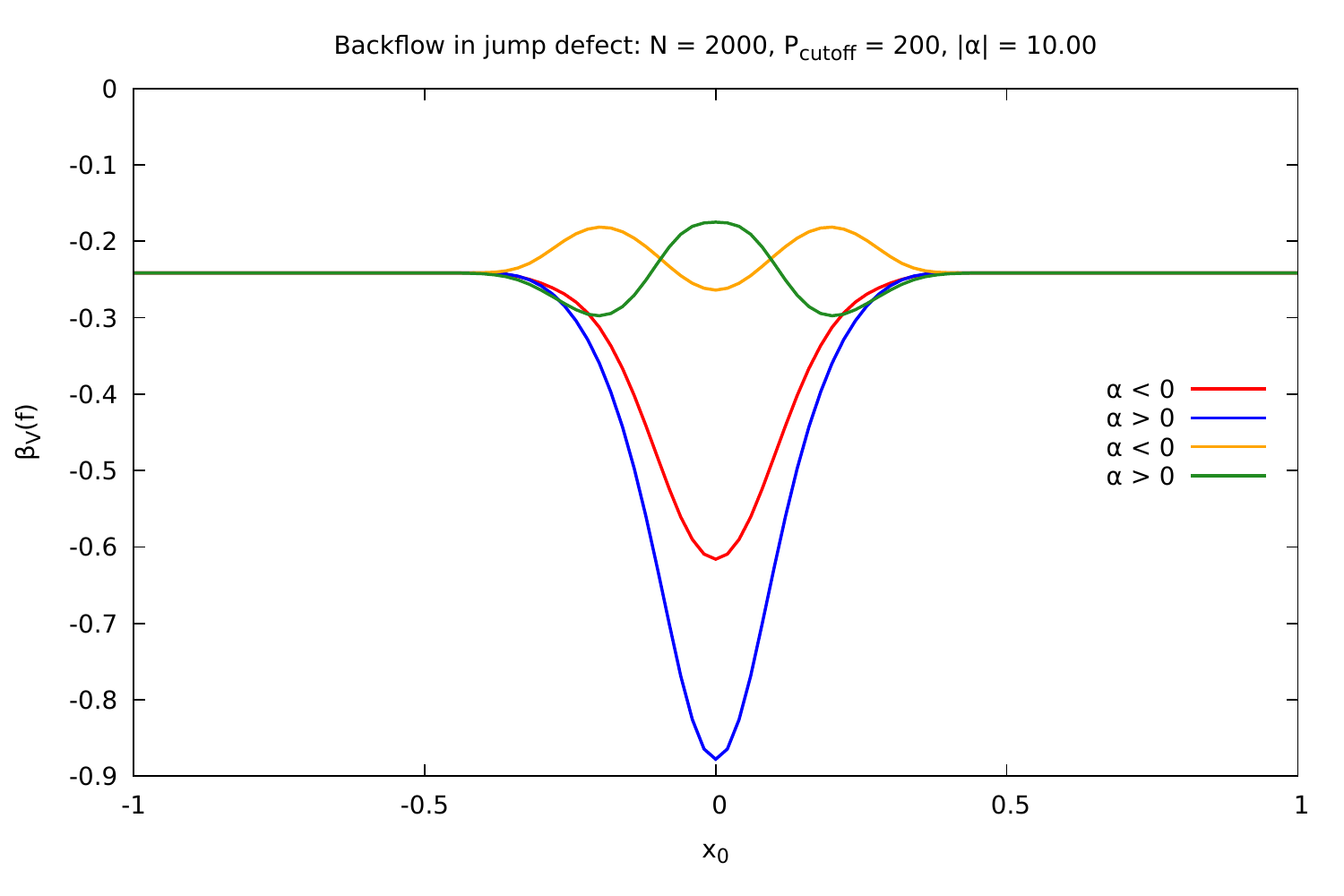} 
 \caption{}
  \end{subfigure}
   \caption{Lowest backflow eigenvalue of the current operator. \!\textcolor{red}{\textbf{Red}}/\textcolor{blue}{\textbf{blue}} refer to the non-conserved probability current. \textcolor{orange}{\textbf{Yellow}}/\textcolor{OliveGreen}{\textbf{green}} refer to the conserved one. (a) $|\alpha|  = 9.0$. (b) $|\alpha|  = 10.0$.} \label{fig:alpha9and10}
  \end{figure}

\begin{figure}[!htb]
  \begin{subfigure}{\linewidth}
  \includegraphics[width=0.95\linewidth]{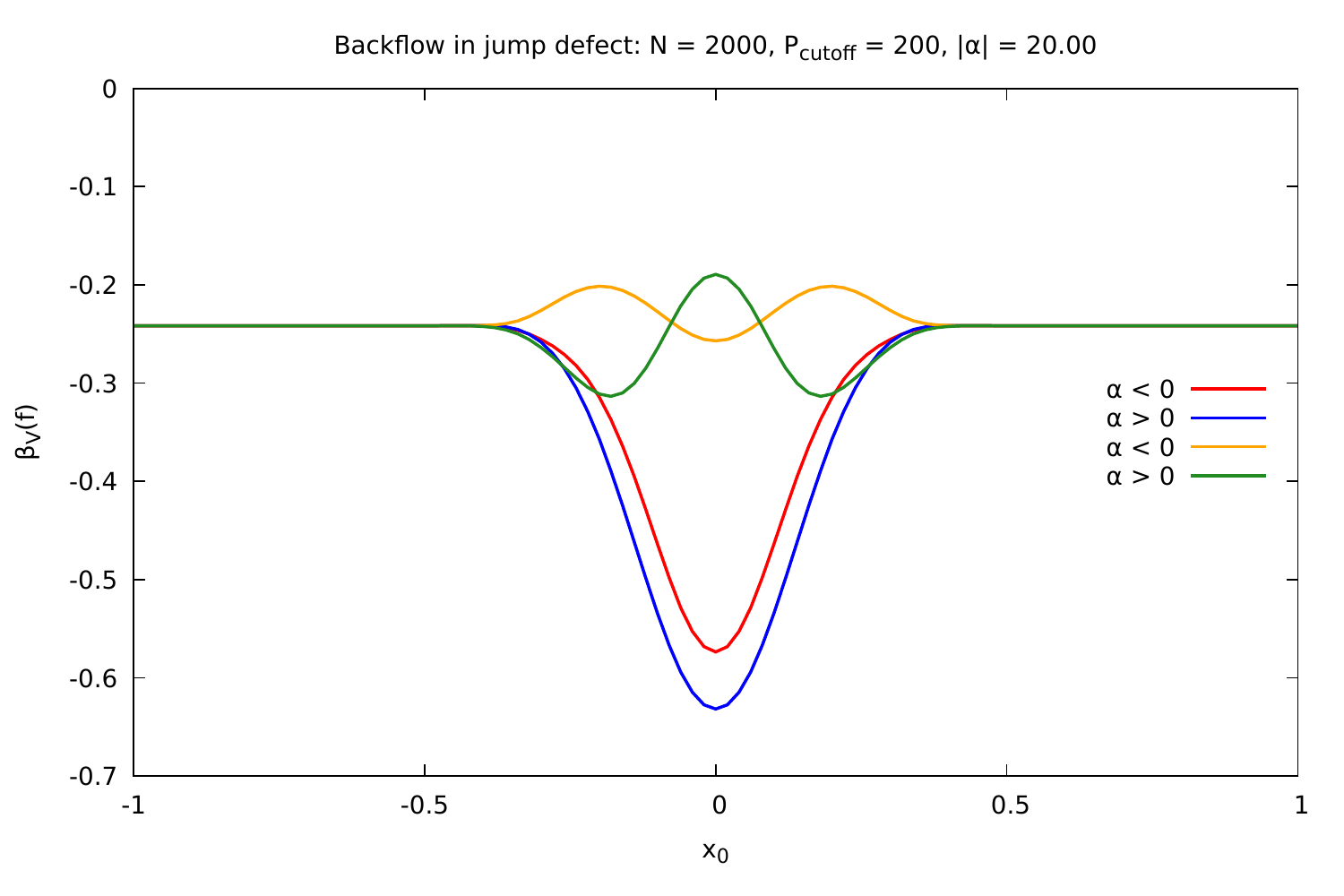}\caption{}
  \end{subfigure}
  \begin{subfigure}{\linewidth}
  \includegraphics[width=0.95\linewidth]{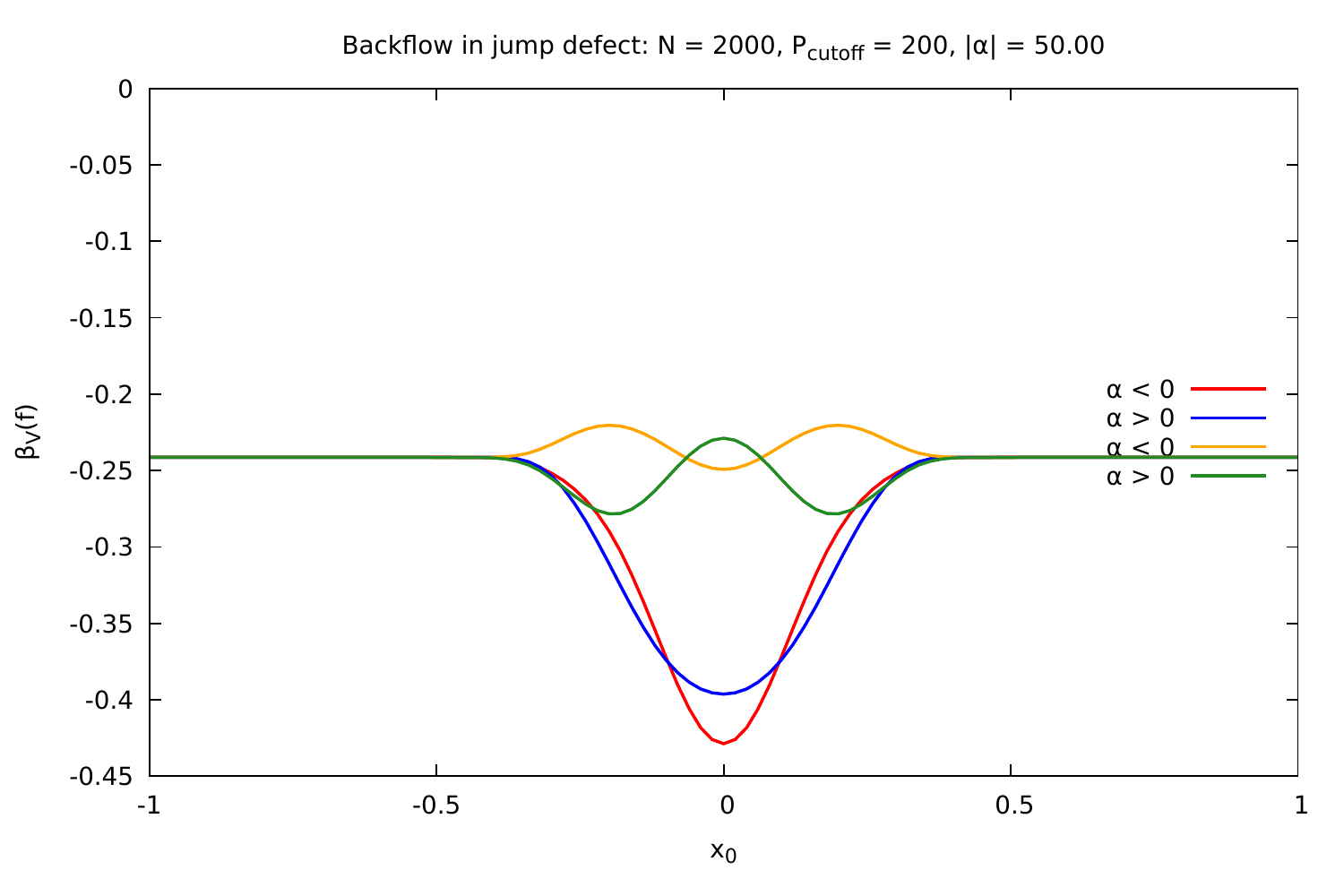}\caption{}
  \end{subfigure}
   \caption{Lowest backflow eigenvalue of the current operator. \!\textcolor{red}{\textbf{Red}}/\textcolor{blue}{\textbf{blue}} refer to the non-conserved probability current. \textcolor{orange}{\textbf{Yellow}}/\textcolor{OliveGreen}{\textbf{green}} refer to the conserved one. (a) $|\alpha|  = 20$. (b) $|\alpha|  = 50$.} \label{fig:alpha20and50}
  \end{figure}

\begin{figure}[!htb]
  \begin{subfigure}{\linewidth}
  \includegraphics[width=0.95\linewidth]{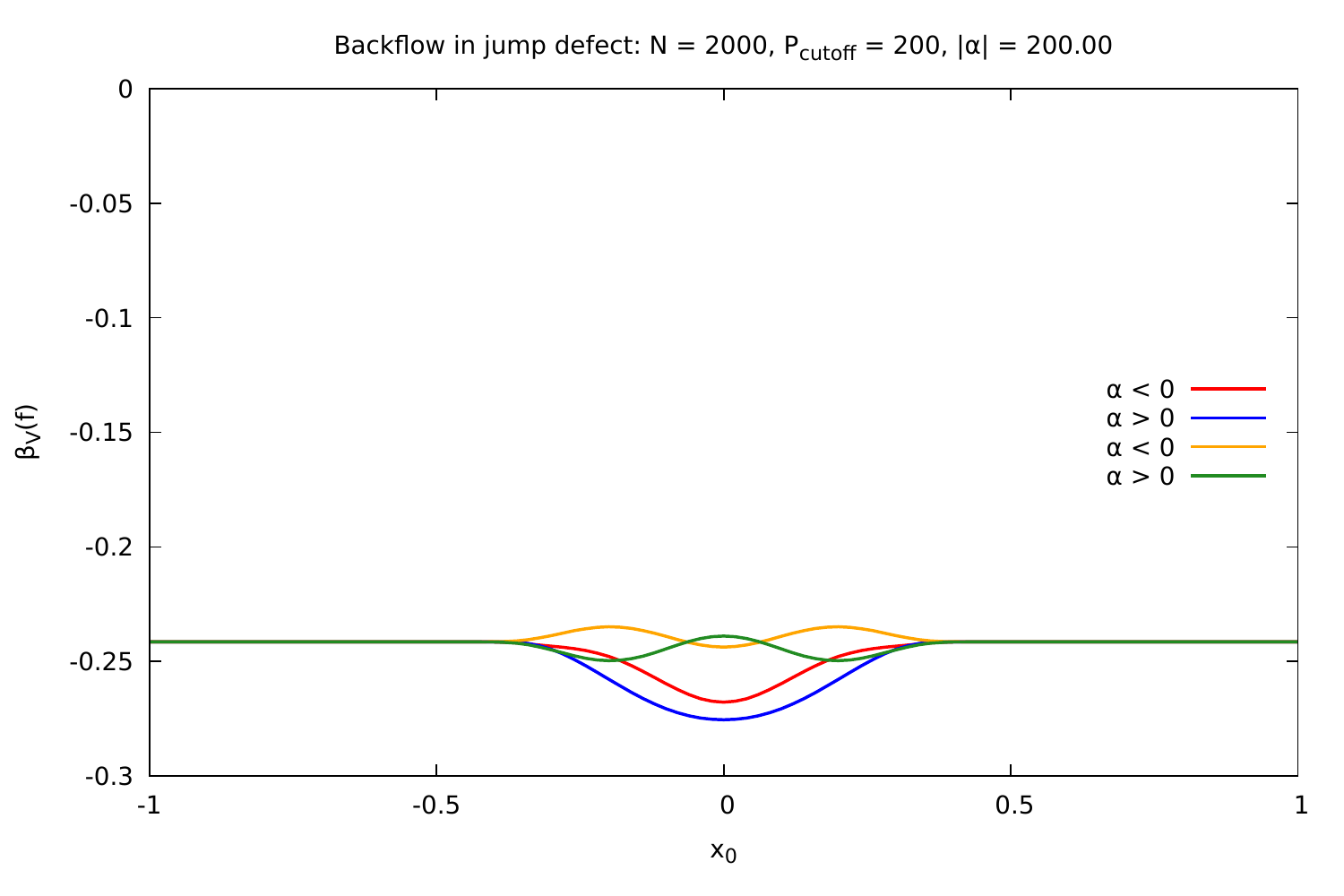}\caption{}
  \end{subfigure}
  \begin{subfigure}{\linewidth}
  \includegraphics[width=0.95\linewidth]{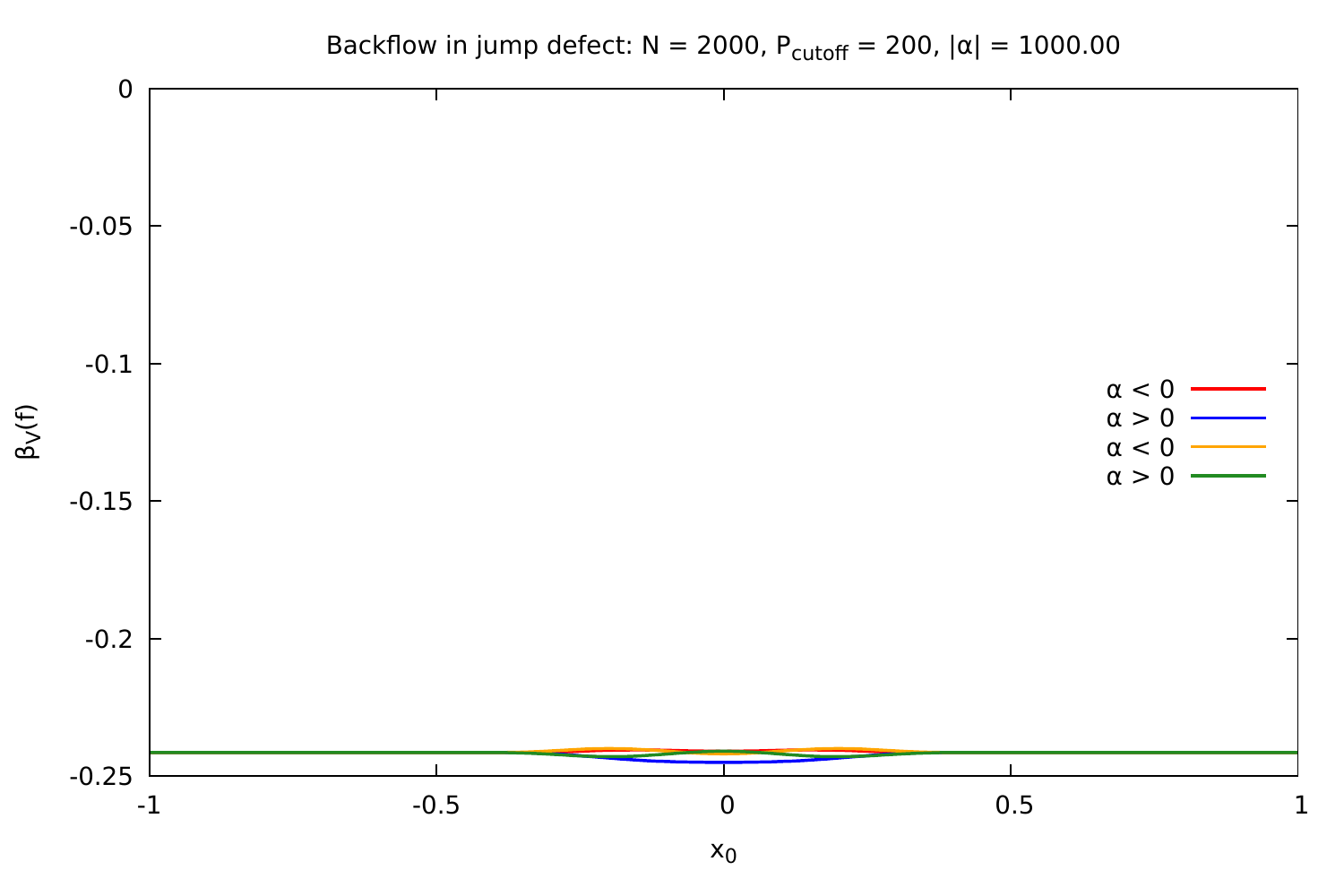}\caption{}
  \end{subfigure}
   \caption{Lowest backflow eigenvalue of the current operator. \!\textcolor{red}{\textbf{Red}}/\textcolor{blue}{\textbf{blue}} refer to the non-conserved probability current. \textcolor{orange}{\textbf{Yellow}}/\textcolor{OliveGreen}{\textbf{green}} refer to the conserved one. (a) $|\alpha|  = 200$. (b) $|\alpha|  = 1000$.} \label{fig:alpha200and1000}
  \end{figure}

Additionally to the two-dimensional plots, we have varied the parameters to display a three-dimensional picture of how the lowest backflow eigenvalue $\beta_V(f)$ is affected in the presence of the jump-defect. For comparison, we have plotted both cases: $\beta_V(f)$ for the non-conserved probability current operator, figure~\ref{fig:landscapenonconserved1} and figure~\ref{fig:landscapenonconserved2}, and for the conserved probability current operator, figure~\ref{fig:landscapeconserved1} and figure~\ref{fig:landscapeconserved2}. All these can be found in the Appendix.

\end{subsection}

\clearpage

\end{section}

\newpage

\section{Concluding remarks}\label{sec:remarks}

The quantum energy inequalities, both in quantum mechanics and quantum field theory, are conditions upon how much spatially and temporally averaged energy densities and fluxes are bounded below. Inequalities for free theories were explored before, but there is no general result for interacting models even in the quantum mechanics theory. The case of the probability flux inequality is also a quantum inequality and it is called quantum backflow. For the interaction-free situation, that the backflow effect is limited in space can be seen from the sharp G\r{a}rding inequalities as shown in \cite{Fewster}. The extension of backflow to scattering situations in short-range potentials was established in \cite{Henning}, where the interacting potential function is assumed to be in the $L^{1+}(\mathbb{R})$-class.

The present work focused on the spatial average probability current operator in the case of a quantum mechanical system in the presence of a local impurity, the discontinuous and purely transmitting jump-defect. As a similar interesting case, we also presented the $\delta$-defect results for comparison. The lowest averaged backflow eigenvalue in the presence of a jump-defect was shown to be spatially constrained even though it has no explicit potential function to be classified in the $L^{1+}(\mathbb{R})$-class. In particular, there is a lower bound on the spatial extent of the backflow effect for both the non-conserved and conserved current operator. Whilst the maximum amount of backflow in the presence of a $\delta$-defect was also found to be bounded, a striking difference between the $\delta$-defect case and the jump-defect is that the lowest eigenvalue can get increasingly negative as the parameter $\lambda \rightarrow \pm \infty $ on the left of a $\delta$-defect and tends to zero on the right of it, but, in the case of the jump-defect, the lowest eigenvalue on regions sufficiently far from its location, both on the left and on the right, is simply equal to the interaction-free situation with the asymptotic backflow constant given by $\beta_0(f) \approx -0.241$. An interesting new observation is the existence of a maximum of the lowest backflow eigenvalue $\beta_V(f)$ that seems to peak for the defect parameter $\lambda=-1/2$ in the $\delta$-defect case. The existence of maxima is even more evident in the jump-defect case where the number of stationary points varies with changes of the defect parameter $\alpha$. These may be related to the presence of bound states, and it is worth seeking an explanation behind the existence and the location of maxima.  

In a field theory of physical interest, we want to keep not only the energy conserved but also the momentum. Equivalently in the quantum mechanics setup, we want to keep the probability current conserved together with the energy and the total probability. In our non-relativistic massive case in the presence of a defect, an energy inequality is not the same as the backflow inequality, but they may be intimately related. While the energy is conserved even in the presence of interaction, the conservation of probability current in quantum mechanics is easily violated by an interaction which breaks the translational symmetry. The jump-defect, in contrast to the $\delta$-defect, which breaks the conservation of the probability current, allows a fixing term, a contribution purely from the defect, to be added in order to maintain its conservation. In this sense, it might be a good model for investigating the relation between these different quantum inequalities.

\section*{Acknowledgements} 
I want to thank Ed Corrigan for proposing the jump-case and for fruitful discussions. I am also thankful to Henning Bostelmann for help with his original code and for discussions, to Petrus Yuri for help with FORTRAN, and to members of the University of York IT Services for assistance using the Viking computing cluster. I thank the University of York Graduate Research School and Department of Mathematics for their financial support.

\newpage
\section*{Appendix: Three-dimensional plot}\label{appendix}

Here you find three-dimensional plots displaying the lowest eigenvalue $\beta_V(f)$ of the corresponding probability current operator as the defect parameter and the position of measurement $x_0$, which is the center of the averaging Gaussian function $f$, change. For the jump-defect, both the non-conserved and the conserved probability current were considered. In each case, we have plotted a version that runs over a large range of the defect parameter and another one that runs over a smaller range for capturing some local details. For the $\delta$-defect case, this was not necessary and we plotted $\beta_V(f)$ against only one single range of the defect parameter.

\begin{sidewaysfigure}[ht]
    \includegraphics[width=1.0\textwidth]{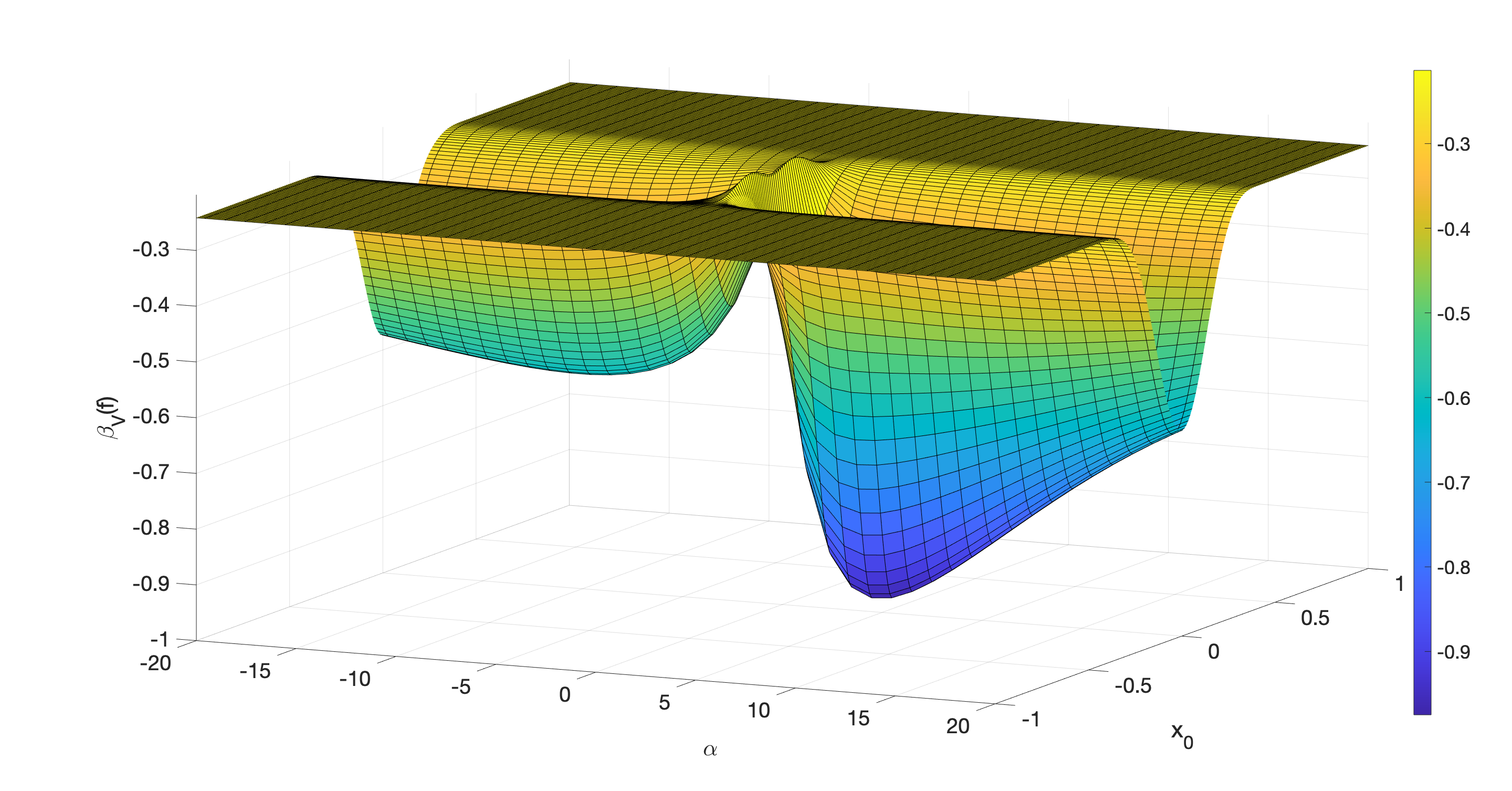}
    \caption{Probability current lowest eigenvalue, $P_{\textrm{cutoff}}=200$, $N=2000$.}
    \label{fig:landscapenonconserved2}
\end{sidewaysfigure}

\begin{sidewaysfigure}[!htb]
    \includegraphics[width=1.0\textwidth]{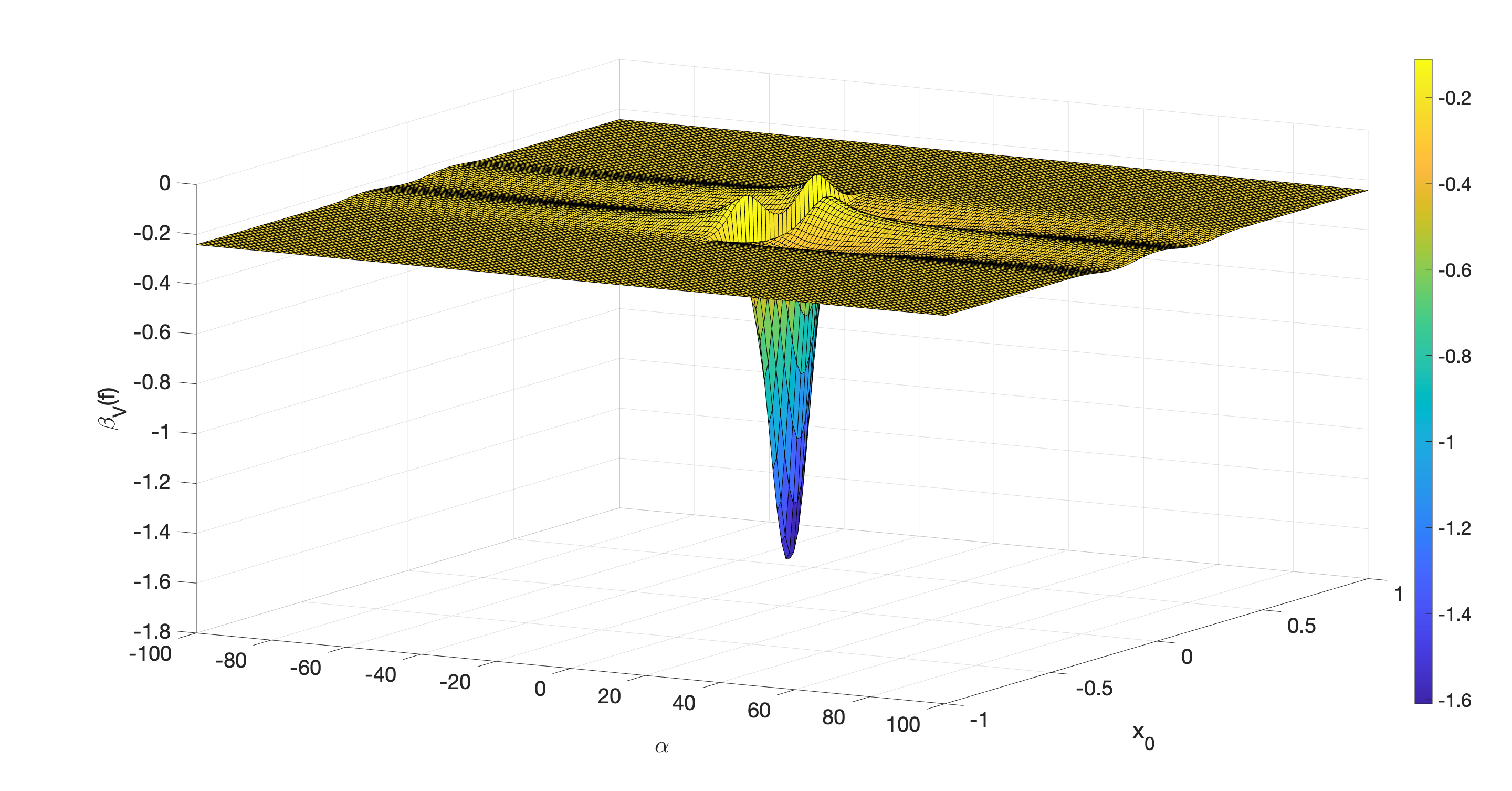}
    \caption{Conserved probability current lowest eigenvalue, $P_{\textrm{cutoff}}=200$, $N=2000$.}
    \label{fig:landscapeconserved1}
\end{sidewaysfigure}

\begin{sidewaysfigure}[!htb]
    \includegraphics[width=1.0\textwidth]{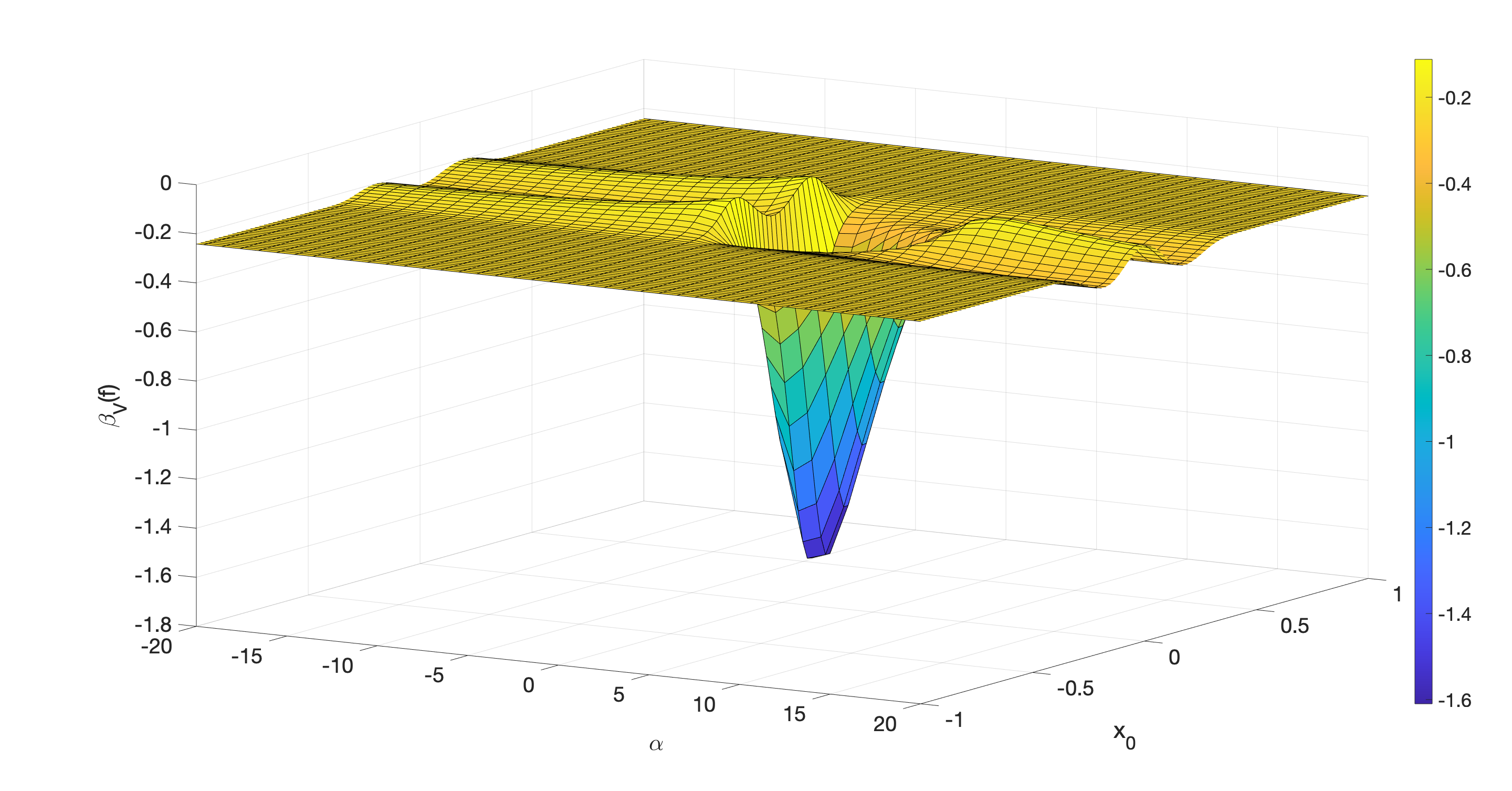}
    \caption{Conserved probability current lowest eigenvalue, $P_{\textrm{cutoff}}=200$, $N=2000$.}
    \label{fig:landscapeconserved2}
\end{sidewaysfigure}

\clearpage

\newpage
\bibliography{References}

\begin{thebibliography}{99}

\bibitem{Bert}
Schroer, B., 2010. Localization and the interface between quantum mechanics, quantum field theory and quantum gravity I: the two antagonistic localizations and their asymptotic compatibility. Studies in History and Philosophy of Science Part B: Studies in History and Philosophy of Modern Physics, 41(2), pp.104-127.

\bibitem{Ford1978}
Ford, L.H., 1978. Quantum coherence effects and the second law of thermodynamics. Proceedings of the Royal Society of London. A. Mathematical and Physical Sciences, 364(1717), pp.227-236.

\bibitem{FewsterQFT}
 Fewster, C.J., 2012. Lectures on quantum energy inequalities. arXiv preprint arXiv:1208.5399. 
 
 \bibitem{Allcock}
 Allcock, G.R., 1969. The time of arrival in quantum mechanics I. Formal considerations. Annals of physics, 53(2), pp.253-285.
 
\bibitem{EpsteinGlaser}
Epstein, H., Glaser, V. and Jaffe, A., 1965. Nonpositivity of the energy density in quantized field theories. Il Nuovo Cimento (1955-1965), 36(3), pp.1016-1022.
 
 \bibitem{Bracken2}
 Melloy, G.F. and Bracken, A.J., 1998. Probability backflow for a Dirac particle. Foundations of physics, 28(3), pp.505-514.

\bibitem{Henning}
Bostelmann, H., Cadamuro, D. and Lechner, G., 2017. Quantum backflow and scattering. Physical Review A, 96(1), p.012112.

 \bibitem{Delfino}
 Delfino, G., Mussardo, G. and Simonetti, P., 1994. Scattering theory and correlation functions in statistical models with a line of defect. Nuclear Physics B, 432(3), pp.518-550.

 \bibitem{Konik1999}
Konik, R. and LeClair, A., 1999. Purely transmitting defect field theories. Nuclear Physics B, 538(3), pp.587-611.


\bibitem{Teller1933}
 P\"{o}schl, G. and Teller, E., 1933. Bemerkungen zur Quantenmechanik des anharmonischen Oszillators. Zeitschrift f\"{u}r Physik, 83(3-4), pp.143-151.

\bibitem{Fewster}
Eveson, S.P., Fewster, C.J. and Verch, R., 2005, February. Quantum inequalities in quantum mechanics. In Annales Henri Poincar\'e (Vol. 6, No. 1, pp. 1-30). Birkh{\"a}user-Verlag.




\bibitem{Bracken}
Bracken, A.J. and Melloy, G.F., 1994. Probability backflow and a new dimensionless quantum number. Journal of Physics A: Mathematical and General, 27(6), p.2197.


\bibitem{Berry}
Berry, M.V., 2010. Quantum backflow, negative kinetic energy, and optical retro-propagation. Journal of Physics A: Mathematical and Theoretical, 43(41), p.415302

\bibitem{Goussev}
Goussev, A., 2020. Probability backflow for correlated quantum states. arXiv preprint arXiv:2002.03364.


\bibitem{Faddeev}
Faddeev, L.D., 1964. Properties of the S-matrix of the one-dimensional Schrodinger equation. Trudy Matematicheskogo Instituta imeni VA Steklova, 73, pp.314-336.

\bibitem{Yafaev:analytic}
Yafaev, D.R., 2010. Mathematical scattering theory: analytic theory (No. 158). American Mathematical Soc..

\bibitem{Deift}
Deift, P. and Trubowitz, E., 1979. Inverse scattering on the line. Communications on Pure and Applied Mathematics, 32(2), pp.121-251.








\bibitem{ReedSimon}
Reed, M. and Simon, B., 1975. II: Fourier Analysis, Self-Adjointness (Vol. 2). Elsevier.



\bibitem{Mintchev2002}
 Mintchev, M., Ragoucy, E. and Sorba, P., 2002. Scattering in the presence of a reflecting and transmitting impurity. Physics Letters B, 547(3-4), pp.313-320.

\bibitem{Ed2003}
 Bowcock, P., Corrigan, E. and Zambon, C., 2004. Classically integrable field theories with defects. International Journal of Modern Physics A, 19(supp02), pp.82-91.
 
 \bibitem{Ed2006}
 Corrigan, E. and Zambon, C., 2006. Jump-defects in the nonlinear Schrodinger model and other non-relativistic field theories. Nonlinearity, 19(6), p.1447.

\bibitem{Caudrelier2008}
Caudrelier, V., 2008. On a systematic approach to defects in classical integrable field theories. International Journal of Geometric Methods in Modern Physics, 5(07), pp.1085-1108.

\bibitem{Ed2009}
 Corrigan, E. and Zambon, C., 2009. A new class of integrable defects. Journal of Physics A: Mathematical and Theoretical, 42(47), p.475203.
 
\bibitem{Goodman} 
Goodman, R.H., Holmes, P.J. and Weinstein, M.I., 2002. Interaction of sine-Gordon kinks with defects: phase space transport in a two-mode model. Physica D: Nonlinear Phenomena, 161(1-2), pp.21-44. 
 
\bibitem{Rankine}
Rankine, W.J.M., 1870. XV. On the thermodynamic theory of waves of finite longitudinal disturbance. Philosophical Transactions of the Royal Society of London, (160), pp.277-288.


 
 \bibitem{Lamb1974}
Lamb Jr, G.L., 1974.  B\"{a}cklund transformations for certain nonlinear evolution equations. Journal of Mathematical Physics, 15(12), pp.2157-2165.




\bibitem{Penz}
Penz, M., Grubl, G., Kreidl, S. and Wagner, P., 2005. A new approach to quantum backflow. Journal of Physics A: Mathematical and General, 39(2), p.423.

\bibitem{Halliwell}
Yearsley, J.M., Halliwell, J.J., Hartshorn, R. and Whitby, A., 2012. Analytical examples, measurement models, and classical limit of quantum backflow. Physical Review A, 86(4), p.042116.



\bibitem{quadpack}
Piessens, R., de Doncker-Kapenga, E., \"{U}berhuber, C.W. and Kahaner, D.K., 2012. Quadpack: a subroutine package for automatic integration (Vol. 1). Springer Science \& Business Media.


\bibitem{eispack}
Smith, B.T., Boyle, J.M., Garbow, B.S., Ikebe, Y., Klema, V.C. and Moler, C.B., 2013. Matrix eigensystem routines-EISPACK guide (Vol. 6). Springer. 







 



 
 
 
 
 
 
 


 

\end{thebibliography}
\bibliographystyle{unsrt}

\end{document}